\documentclass[journal]{new-aiaa}
\usepackage[utf8]{inputenc}
\usepackage{textcomp}

\usepackage{graphicx,tabularray}
\usepackage{amsmath}
\usepackage{siunitx}
\usepackage{longtable,tabularx}
\usepackage{layouts}
\usepackage{subcaption}
\usepackage{pdfpages}
\usepackage{multirow}
\usepackage{threeparttable}
\usepackage{array} 

\title{Role of Duty Cycle in Burst-modulated Synthetic Jet Flow Control}

\author{Adnan Machado\footnote{MASc, Mechanical Engineering. adnan.machado@mail.utoronto.ca}, Ali Shirinzad\footnote{PhD Candidate}, Pierre Sullivan\footnote{Associate Fellow}}
\affil{Mechanical and Industrial Engineering, University of Toronto, Toronto, Ontario, M5S 3G8}
\author{Kecheng Xu\footnote{PhD, Aerospace Engineering}}
\affil{Institute for Aerospace Studies, University of Toronto, Toronto, Ontario, M3H 5T6}

\begin{document}

\maketitle

\begin{abstract}
The effect of duty cycle (DC) and blowing ratio on synthetic jet flow control over a stalled NACA 0025 airfoil at $\mathrm{Re}_c=10^5$ was investigated experimentally. A finite-span microblower array operating with burst modulation was tested across a wide range of control parameters to assess aerodynamic performance, power consumption, and flow stability. Flow reattachment was achieved once a threshold momentum coefficient was met via increasing either the DC or blowing ratio. Control effectiveness increased sharply upon reattachment, with additional momentum providing incremental improvements in lift, spanwise control, and flow stability, though these effects eventually saturated. Substantial lift improvements are observed at DCs as low as \SI{5}{\percent}, indicating that brief, high-momentum bursts were the most power-efficient for achieving reattachment. However, flow stability was reduced at low DCs due to the inconsistent streamwise dissipation of spanwise vortices responsible for flow control. Higher DC control strategies resulted in more consistent boundary layer control. These results provide a framework for selecting control strategies that balance aerodynamic performance and stability with power efficiency.
\end{abstract}
\newpage
\section*{Nomenclature}
{\renewcommand\arraystretch{1.0}
\noindent\begin{longtable*}{@{}l @{\quad=\quad} l@{}}
$A_f$  & Projected control area [\unit{\metre\squared}] \\
$b$ & Wing span [\unit{\milli\metre}]\\
$c$ & Airfoil chord length [\unit{\milli\metre}]\\
$C_B$ & Blowing ratio \\
$C_L$ & Lift coefficient \\
$C_{LO}$ & Baseline lift coefficient \\
$c_p$ & Instantaneous pressure coefficient \\
$C_p$ & Mean pressure coefficient \\
$C_\mu$ & Momentum coefficient \\
$f_c$ & Carrier frequency [\unit{\hertz}]\\
$f_m$ & Modulation frequency [\unit{\hertz}]\\
$F^+$ & Reduced SJA frequency \\
$\overline{I_j}$ & Time-averaged jet momentum [\unit{\newton\second}]\\
$P$ & Power consumption [W]\\
$Q$ & Second invariant of the velocity gradient tensor \\
$R$ & Resistance [$\Omega$] \\
$\mathrm{Re}_c$ & Chord-based Reynolds number \\
$S$ & Strain-rate tensor \\
$T$ & Actuation period [\unit{\second}]\\
$u, v$ & Instantaneous velocity in the streamwise and transverse direction, [\unit[per-mode=symbol]{\metre\per\second}] \\
$\bar{u}$, $\bar{v}$ & Time-averaged velocity in the streamwise and transverse direction [\unit[per-mode=symbol]{\metre\per\second}] \\
$\tilde{u}, \tilde{v}$ & Phase-averaged velocity in the streamwise and transverse direction, [\unit[per-mode=symbol]{\metre\per\second}] \\
$u', v'$ & Velocity fluctuation in the streamwise and transverse direction, [\unit[per-mode=symbol]{\metre\per\second}] \\
$U_\infty$ & Freestream velocity [\unit[per-mode=symbol]{\metre\per\second}]\\
$\overline{U_j}$ & Time-averaged centerline jet velocity during expulsion [\unit[per-mode=symbol]{\metre\per\second}]\\
$V$ & Voltage [V] \\
$|\overline{\mathbf{V}}|$ & Time-averaged velocity magnitude [\unit[per-mode=symbol]{\metre\per\second}]\\
$x$, $y$, $z$ & Chordwise, transverse, and spanwise coordinate [\unit{\milli\metre}]\\
$\alpha$ & Angle of attack [\unit{\degree}] \\
$\eta$ & Lift to power efficiency [\unit{\per\watt}]\\
$\rho_\infty$ & Freestream fluid density [\unit[per-mode=symbol]{\kilo\gram\per\metre\cubed}]\\
$\rho(A,B)$ & Pearson correlation coefficient\\
$\tau$ & Burst signal active duration [s]\\
$\phi$ & Phase angle of modulated actuation cycle [\unit{\degree}]\\
$\omega_z$ & Out-of-plane vorticity [\unit{\per\second}]\\
\multicolumn{2}{@{}l}{Acronyms}\\
AFC & Active Flow Control \\
DC & Duty Cycle \\
FOV & Field of View \\
HWA & Hot-wire anemometry \\
NACA & National Advisory Committee for Aeronautics \\
NREL & National Renewable Energy Laboratory \\
PIV & Particle Image Velocimetry \\
SJA & Synthetic Jet Actuator \\
\end{longtable*}}

\section{Introduction}
Flow control plays a critical role in enhancing the performance and efficiency of aircraft, especially those operating at low Reynolds numbers. Aerodynamic stall due to flow separation results in a drastic loss of lift and an increase in drag, constraining the performance of airfoils. These challenges typically arise in low-speed, high-altitude aircraft used for surveying and stealth, gliders, electric aircraft, and wind turbine blades. Flow control techniques can modify the flow around an airfoil, reattaching separated flows during post-stall conditions, thereby expanding the airfoil's operational envelope.

Flow control techniques can be broadly classified into passive and active methods. Passive techniques involve modifying the flow without adding energy, for example, modifying the wing shape, adding vortex generators~\cite{Lin2002}, or incorporating roughness elements~\cite{Zhou2012}. While passive flow control can enhance aerodynamic performance, these methods often incur parasitic drag in portions of the operational envelope where they are not needed~\cite{Taylor2015}. In contrast, active flow control (AFC) techniques can be selectively employed under specific conditions, minimizing adverse effects when inactive. Despite their advantages, AFC systems tend to increase aircraft weight, which is especially problematic for nano- and micro-air vehicles, where weight and volume limitations are stringent~\cite{Greenblatt2022}. Furthermore, AFC systems draw substantial electrical power, a significant concern for electric aircraft~\cite{Greenblatt2022,Viswanathan2022}. Traditional continuous blowing methods, for instance, require plumbing systems that add weight and complexity~\cite{Greenblatt2000}. In contrast, AFC using synthetic jet actuators (SJAs) does not require plumbing, making them an ideal choice for aircraft flow control.

Typically utilizing plasma actuators or piezoelectric diaphragms, SJAs are zero-net-mass-flux devices that cyclically ingest and expel fluid, forming a time-averaged jet that adds momentum to the flow~\cite{Smith1998,Glezer2002}. Furthermore, synthetic jets have been shown to outperform continuous jets in fluid entrainment and mixing~\cite{Seifert1999,Smith2003,DiCicca2007}. The effectiveness of SJAs in flow control is mainly due to the generation of coherent vortical structures. These vortices, induced by the cyclic actuation of SJAs, are convected downstream along the suction side of the airfoil, entraining high-momentum fluid from the freestream into the separated shear layer. This momentum transfer energizes the shear layer, enabling it to overcome the adverse pressure gradient, thereby facilitating flow reattachment~\cite{Greenblatt2000,Salunkhe2016,Rice2018,Xu2023}. The ability of SJAs to control flow separation through this mechanism makes them particularly effective for enhancing aerodynamic performance. Recent advancements in SJA technology have led to the development of MEMS-based devices, which are lightweight and reliable, making them ideal for flow control applications~\cite{Xu2023}. Additionally, SJAs offer a cost-effective alternative to similar AFC devices~\cite{Taylor2015}. Over the past few decades, SJAs have been extensively researched and have achieved a high technology readiness level of 7 through successful implementation in real flight tests~\cite{Seifert2010,Greenblatt2022}.

Slot-style SJAs have been studied extensively due to their advantages in fluid entrainment~\cite{Toyoda2009}. However, their installation requires significant modifications to the wing surface and often involves large cavities, raising structural concerns and increasing power consumption. This has driven research into arrays of small, circular SJAs for flow control, which also offer low power requirements and minimal noise compared to large-cavity SJAs. For example, a previous study using a large cavity SJA required driving voltages up to 275~$\mathrm{V_{pp}}$~\cite{Goodfellow2013}, while the effects of control were fully saturated at only 20~$\mathrm{V_{pp}}$ using an array of microblowers under similar experimental conditions~\cite{Xu2023}.

Many parametric studies have been conducted to optimize flow reattachment, reduce drag, enhance lift, and maximize the stability of aerodynamic forces. Research has examined various factors, such as the effect of chordwise actuation location~\cite{Amitay2001,Tang2014,Feero2017b}, blowing angle~\cite{Amitay2001,Shuster2005}, and orifice geometry~\cite{Jeyalingam2016,VanBuren2016a}. For practical flight applications, the most easily adjustable parameters that affect control performance are related to the waveform driving the SJAs. For a burst-modulated signal, the three primary waveform parameters that can be adjusted are the modulation frequency, the blowing strength, and the burst duty cycle (DC). Power consumption is affected by both the blowing strength, controlled by the applied voltage, and the DC, which dictates the fraction of the period during which the SJA is activated.

The blowing strength is a critical parameter for SJAs in flow control applications, often quantified by a non-dimensional blowing ratio $C_B = \overline{U_j}/U_\infty$, where $\overline{U_j}$ is the centerline, time-averaged jet velocity at the orifice exit during expulsion, and $U_\infty$ is the freestream velocity. The blowing strength of an SJA is dictated by the amplitude of the SJA's actuation, which is controlled by the voltage amplitude of the input signal. Experimental studies have demonstrated that increasing the blowing ratio can result in reduced drag and increased lift~\cite{Seifert1999,Amitay2001,Goodfellow2013,Feero2017b,Yang2022}. However, the effectiveness eventually saturates, with no additional aerodynamic benefits from further increases in the blowing ratio. This saturation point corresponded to the transition from a laminar separation bubble to a fully reattached flow~\cite{Feero2017a}. Prior flow visualizations revealed that an increase in the blowing ratio leads to a thinning of the boundary layer, a decrease in wake width, and a downward shift of the wake~\cite{Goodfellow2013}. These desirable flow characteristics observed at higher blowing ratios are attributed to the postponed dissipation of the coherent vortices induced by the SJA, which enhances momentum transport into the boundary layer~\cite{Liu2022,Yang2022}. It is important to note that the majority of findings in these studies have been based on data collected at the midspan. A recent study revealed that flow over an airfoil controlled by SJAs is highly three-dimensional, with control efficacy degrading significantly even at moderate distances from the midspan, though still within the extent of the SJA array~\cite{Machado2024a}. Furthermore, \citet{Feero2017a} showed that, while control effects may saturate at midspan, further increases in the blowing ratio can improve spanwise control authority, highlighting the need for comprehensive three-dimensional flow analysis.

For flow control applications, SJAs are often driven by a burst-modulated signal, allowing the SJA to operate at its optimal frequency while targeting global instabilities in the flow associated with significantly lower frequencies. The modulation frequency, $f_m$ is often represented by the nondimensional reduced frequency, \begin{equation}F^+=\frac{f_mc}{U_\infty},\end{equation} where $c$ is the chord length. This parameter is defined similarly to the Strouhal number and is used to characterize the actuation unsteadiness relative to the convective timescale. Burst modulation has proven to be an effective technique in flow control, as demonstrated by various experimental studies~\cite{Amitay2002b,Glezer2005,feero2015a,Feero2017a,Feero2017b,Borghi2017,An2017,Rice2018,Rice2021,Yang2022,Kim2022,Xu2023,Xu2025effect}. Additionally, it offers significant advantages over continuous actuation; for example, \citet{Amitay2002b} showed that burst modulation, when tuned to the airfoil's natural frequency, increased the lift coefficient by \SI{400}{\percent} compared to continuous high-frequency actuation while operating at \SI{25}{\percent} of the momentum coefficient. Similarly, \citet{Borghi2017} showed that burst modulation of plasma SJAs was significantly more effective in recovering stall than continuous actuation. \citet{An2017} demonstrated successful temporary flow reattachment following a four-pulse burst sequence from an SJA. Additionally, this technique has the benefit of reduced power consumption due to the SJA only being powered for a fraction of the signal's period.

The DC of the burst-modulated signal is another critical parameter in optimizing SJA performance and power consumption. Low-DCs significantly reduce power input, as the actuator operates for only a fraction of the signal's period. In a parametric study, \citet{Abdolahipour2021} found that high-frequency synthetic jets achieve maximum mean exit velocities at low DCs, which is desirable for flow control applications. The study also revealed that high DCs cause successive vortex rings to converge; at excessively high DCs, they interact and lose coherence. Additionally, \citet{Taylor2015} demonstrated that using \SI{20}{\percent} and \SI{60}{\percent} DCs reduced hysteresis in lift compared to continuous actuation for a dynamically pitching airfoil. Similarly, \citet{Rice2018} improved many aspects of dynamic stall control by using a low-DC, consuming only \SI{35}{\percent} of the power required for continuous actuation. Table~\ref{tab:SJA_studies} provides an overview of the discussed experimental studies utilizing burst modulation in synthetic jet flow control.

\begingroup
\renewcommand{\arraystretch}{1.5} 
\begin{table}
\caption{\label{tab:SJA_studies} Summary of experimental flow control studies with burst-modulated SJA actuation}
\centering
\begin{tblr}{llllllll}
\hline\hline
Authors & DC & $C_\mu \times 10^{-3}$ & $F^+$ & $\alpha$ & $\mathrm{Re}_c\times10^5$ & Geometry \\
\hline
\citet{Amitay2002b} & \SI{25}{\percent} & $4.5$  & 0.27, 1.1, 3.3 & \ang{17.5} & 3.10 & custom airfoil \\
\citet{Taylor2015} & \SI{20}{\percent}, \SI{60}{\percent} & $0.9 $ to $4.4 $ & - & pitching & 2.20 & NREL S809 \\
\citet{Borghi2017} & \SI{50}{\percent} & - & 0.7 & \ang{19}, \ang{26} & 2.20 & NACA 0015 \\
\citet{Feero2017b} & \SI{50}{\percent} & - & 1, 2, 14, 58 & \ang{12} & 1.00 & NACA 0025 \\
\citet{Rice2018,Rice2021} & \SI{35}{\percent} & $12 $ & 25 & \ang{21}, \ang{24} & 3.75 & NREL S817 \\
\citet{Yang2022} & \SI{50}{\percent} & - & 1, 14 & \ang{12} & 1.00 & NACA 0025 \\
\citet{Kim2022} & \SI{50}{\percent} & $2.0 $ & 1, 2, 14, 58 & \ang{12} & 1.00 & NACA 0025 \\
\raggedright\citet{Xu2023} & \SI{50}{\percent} & $2.0 $ & 1.18, 11.76 & \ang{10} & 1.00 & NACA 0025 \\
\citet{Machado2024a, Machado2024b}& \SI{50}{\percent} & $2.0 $ & 1.18, 11.76 & \ang{10} & 1.00 & NACA 0025 \\
\hline\hline
\end{tblr}
\end{table}
\endgroup

The fluidic performance of SJAs is often evaluated using the momentum coefficient, which quantifies the momentum imparted by the SJA relative to the freestream flow. The momentum coefficient for burst-modulated SJAs is not only influenced by the peak-to-peak voltage ($\mathrm{V_{pp}}$) but also by the DC, as highlighted by \citet{Margalit2005}. While previous research has often focused on enhancing momentum addition through increased signal amplitude at a constant DC, the blowing ratio ($C_B$) and momentum coefficient ($C_\mu$) have often been used interchangeably as measures of jet strength. In this study, the momentum coefficient is defined as the ratio of the momentum imparted by the synthetic jet to the incoming flow momentum over the entire jet area and actuation cycle:
\begin{equation}
C_\mu = \frac{\overline{I_j}}{\frac{1}{2}\rho_\infty A_fU_\infty^2},
\end{equation}
where $\rho_\infty$ is the freestream fluid density, $A_f$ is the projected control area for a single jet, and $\overline{I_j}$ is the time-averaged jet momentum given by:
\begin{equation}
    \overline{I_j} = \frac{1}{T}\int_{A_{j}} \int_{0}^{T} \rho_\infty \widetilde{u^2_j} \,dt\,dA
\end{equation}
Here, $A_j$ denotes the jet orifice area, $\widetilde{u_j}$ is the phase-averaged jet momentum, and $T$ is the cycle period. Various definitions of the momentum coefficient appear in the literature, for example, by neglecting spatial variations and assuming a top-hat velocity profile at the orifice exit, or by assuming zero phase-averaged velocity during the non-active segment of the burst cycle. Under the latter assumption, the jet momentum is integrated only over the SJA active duration, $\tau = DC \times T$, leading to a momentum coefficient that scales linearly with DC~\cite{Amitay2001,Steinfurth2018,Berthold2020}. This assumption does not necessarily hold, as will be shown for the actuators considered in the present study. The momentum coefficient may still be scaled linearly, assuming that the velocity profile, including transient dynamics, scales proportionally with DC. In general, the linear scaling of the momentum coefficient must be verified before reliable estimates can be made.

To summarize, in burst-modulated SJAs, the momentum coefficient depends on both the blowing ratio and the DC. With these definitions, the blowing ratio measures the jet's mean velocity relative to the oncoming flow, while the DC is the fraction of the actuation period that the synthetic jet is powered. While prior studies have demonstrated that a threshold momentum coefficient must be met to fully reattach the separated shear layer~\cite{Goodfellow2013, Greenblatt2022}, it remains uncertain whether altering the momentum coefficient by varying the DC has the same effect as varying it via the blowing ratio. This raises an important question: can short, intense jets achieve the same control efficacy as longer, less intense bursts when the time-averaged momentum imparted is equal?

This paper explores the aerodynamic effects of varying the DC and blowing ratio of an SJA array to establish relationships between actuation parameters, flow reattachment mechanisms, and the trade-offs governing aerodynamic performance, flow stability, and power efficiency. Firstly, Section~\ref{characterization} details the characterization of microblowers used in this study. Section~\ref{sec:parametric} investigates how the DC and blowing ratio influence lift recovery, spanwise control authority, and power consumption, aiming to identify effective, power-efficient control strategies. Section~\ref{stability} examines the effect of the DC on flow stability, while Section~\ref{sec:duty_cycle} analyzes the flow structures and mechanisms responsible for flow reattachment and stability. Finally, correlations between single-point pressure measurements and lift coefficients are explored in Section~\ref{sec:rapid_lift_estimation}, providing a method for rapid evaluation of control effectiveness.

\section{Experimental Method}
\begin{figure}
    \centering
    \includegraphics[width=0.6\linewidth]{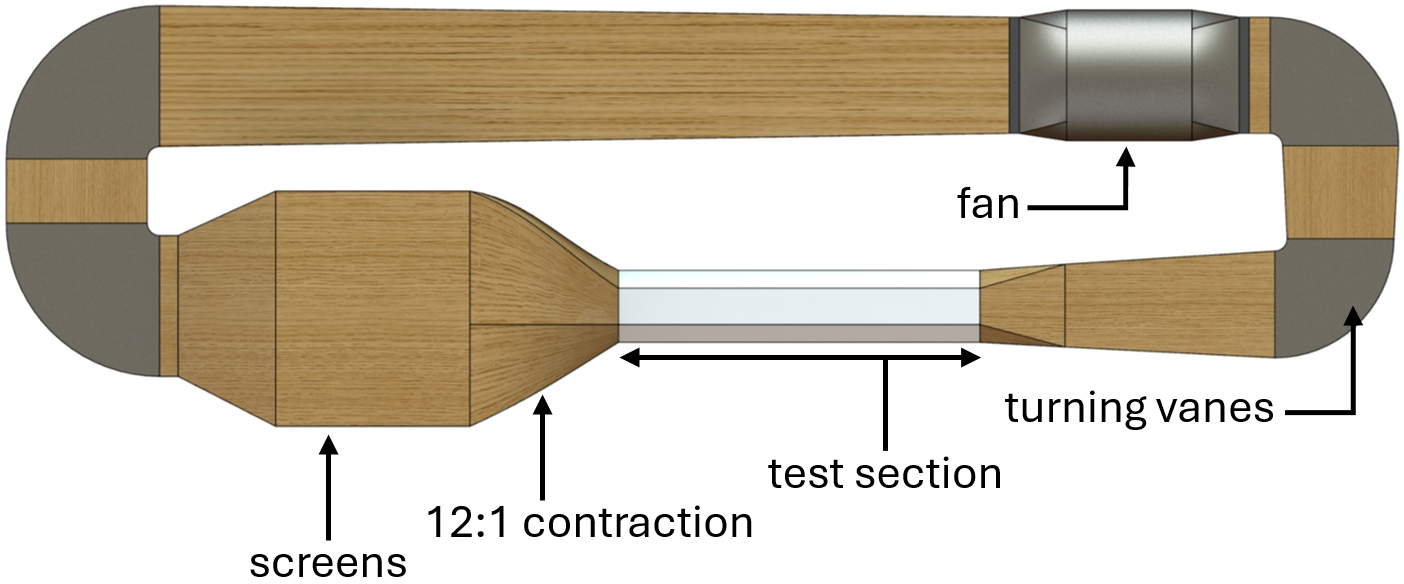}
    \caption{Labelled schematic of the wind tunnel}
    \label{fig:wind tunnel}
\end{figure}

Experiments were conducted in the low-speed recirculating wind tunnel in the Department of Mechanical and Industrial Engineering at the University of Toronto (Figure~\ref{fig:wind tunnel}). The test section has dimensions of \SI{5}{\metre}~$\times$~\SI{0.91}{\metre}~$\times$~\SI{1.22}{\metre} and features acrylic windows on the top and side walls for observation and measurement. The flow passes through seven screens and a 12:1 contraction before entering the test section. The wind tunnel is capable of producing speeds between 3--18~\unit[per-mode = symbol]{\metre\per\second} with a turbulence intensity of less than \SI{1}{\percent}. The freestream velocity was measured with a pitot-static tube at the test section entrance connected to an MKS Baratron 226A differential pressure transducer with a range of \SI{267}{\pascal}. The uncertainty of the freestream velocity is estimated to be less than $\pm \SI{1}{\percent}$. For the experiments conducted, the wind tunnel was operated at a freestream velocity of $U_\infty=5.1$~\unit[per-mode = symbol]{\metre\per\second}, resulting in a chord-based Reynolds number of $\mathrm{Re}_c=10^5$.

\begin{figure}
    \centering
    \includegraphics[width=0.7\linewidth]{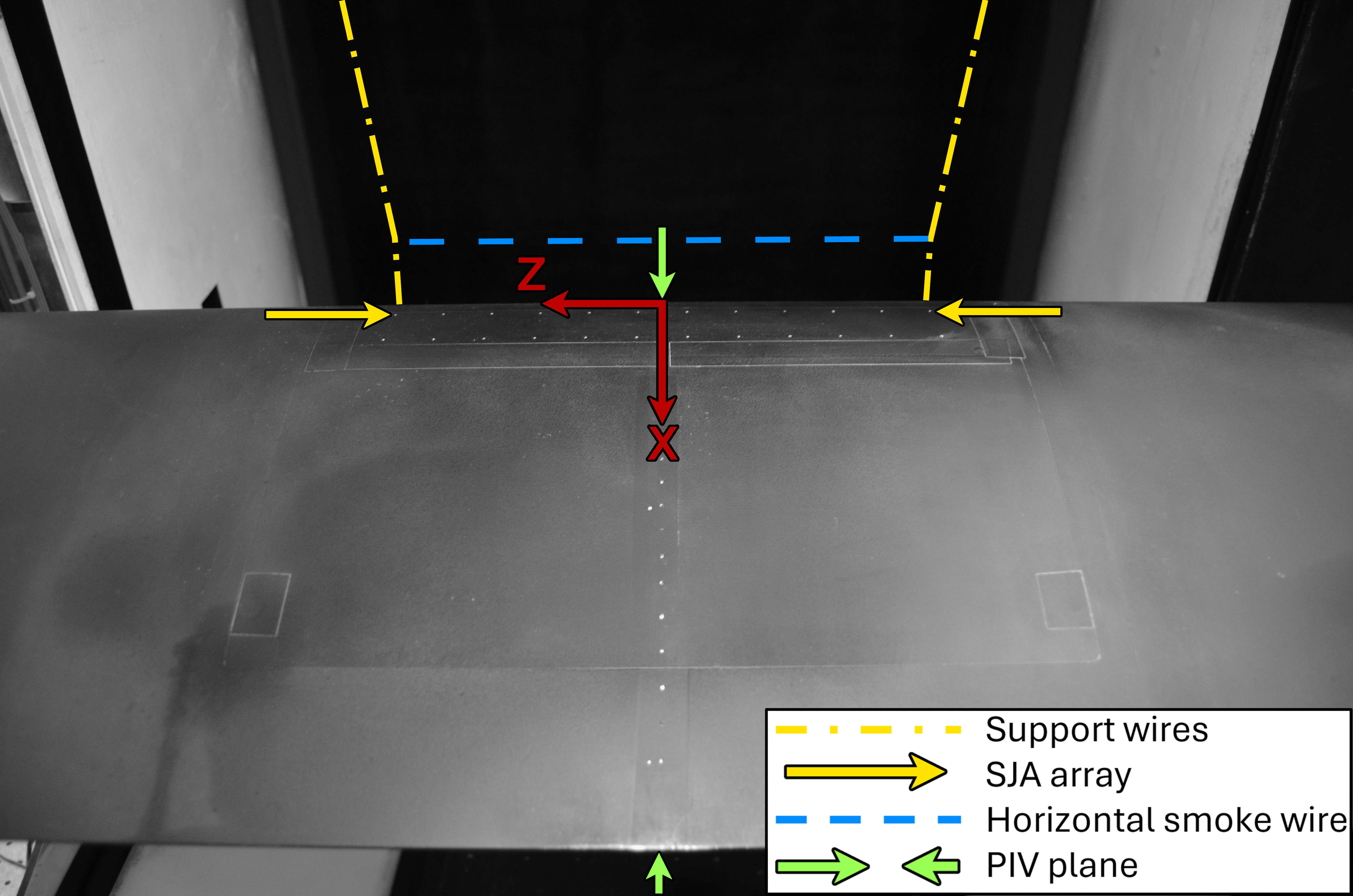}
    \caption{Labelled overhead view of the NACA 0025 airfoil in the test section}
    \label{fig:airfoil}
\end{figure}

A NACA 0025 airfoil was placed in the wind tunnel with the leading edge approximately \SI{40}{\centi\metre} from the test section inlet (Figure~\ref{fig:airfoil}). Additionally, the coordinate axes are defined, with $z=0~\unit{\metre}$ corresponding to the midspan. The aluminum wing has an aspect ratio of approximately 3, with a span of $b=885$~\unit{\milli\metre}, and a chord length of $c=300$~\unit{\milli\metre}. The wing spans the entire width of the test section and features circular end plates, which isolate it from the boundary layer at the wind tunnel walls. With the end plates installed, the spanwise variation in streamwise velocity remained within 6\% of the midspan value, indicating quasi-two-dimensional flow behavior in the baseline condition~\cite{Feero2015b}. It is important to note that this two-dimensionality applies only to the baseline case; the SJA-controlled flows exhibit spanwise variation, which is a focus of the present study. The wing comprises three parts, with a hollow center section to house the sensors and actuators. Near the leading edge of the center section, there is a \SI{317}{\milli\metre}~$\times$~\SI{58}{\milli\metre} rectangular cutout where the microblower array is installed, with a flush \SI{0.8}{\milli\metre} hole for the nozzle of each SJA. The angle of attack was set to $\alpha=\SI{10}{\degree}$, such that the flow separates at approximately \SI{12}{\percent} chord with the specified flow parameters. The experimental parameters are summarized in Table~\ref{tab:experimental parameters}.

\begin{table}
\caption{Summary of experimental parameters\label{tab:experimental parameters}}
\centering
\begin{tabular}{ll}
\hline\hline
        Airfoil profile & NACA 0025 \\
        Chord length & $c=\SI{300}{\milli\metre}$ \\
        Span & $b=885$~\unit{\milli\metre} \\
        Angle of attack & $\alpha=\SI{10}{\degree}$ \\
        Chord-based Reynolds number & $\mathrm{Re}_c=10^5$ \\
        Freestream velocity & $U_\infty=5.1$~\unit[per-mode = symbol]{\metre\per\second} \\
\hline\hline
\end{tabular}
\end{table}

Sixty-four pressure taps along the midspan of the airfoil were connected to a Scanivalve pressure scanner with pneumatic tubing. The surface pressure of the airfoil was measured using an MKS Baratron 226A pressure transducer, featuring a bidirectional range of $\pm26.7$~\unit{\pascal}. For each measurement, 30,000 samples were collected at a sampling rate of \SI{1}{\kilo\hertz}. Surface pressure measurements at each tap were recorded after a 10-second settling period to minimize errors due to sensor response delays. The resulting pressure distribution was used to compute the midspan lift coefficient. The uncertainty due to random error in the mean pressure coefficient was calculated to be $\pm$ \SI{3}{\percent} at the \SI{95}{\percent} confidence level for all measured scenarios and control cases. These uncertainties were propagated through the pressure-integration procedure used to compute lift, resulting in an uncertainty in the lift coefficient due to random error of $\pm$ \SI{4}{\percent}. To account for potential bias errors in the measurement, the total uncertainty of the lift coefficient is estimated to be $\pm$ \SI{5}{\percent}.

\begin{figure}
    \centering
    \includegraphics[width=0.5\linewidth]{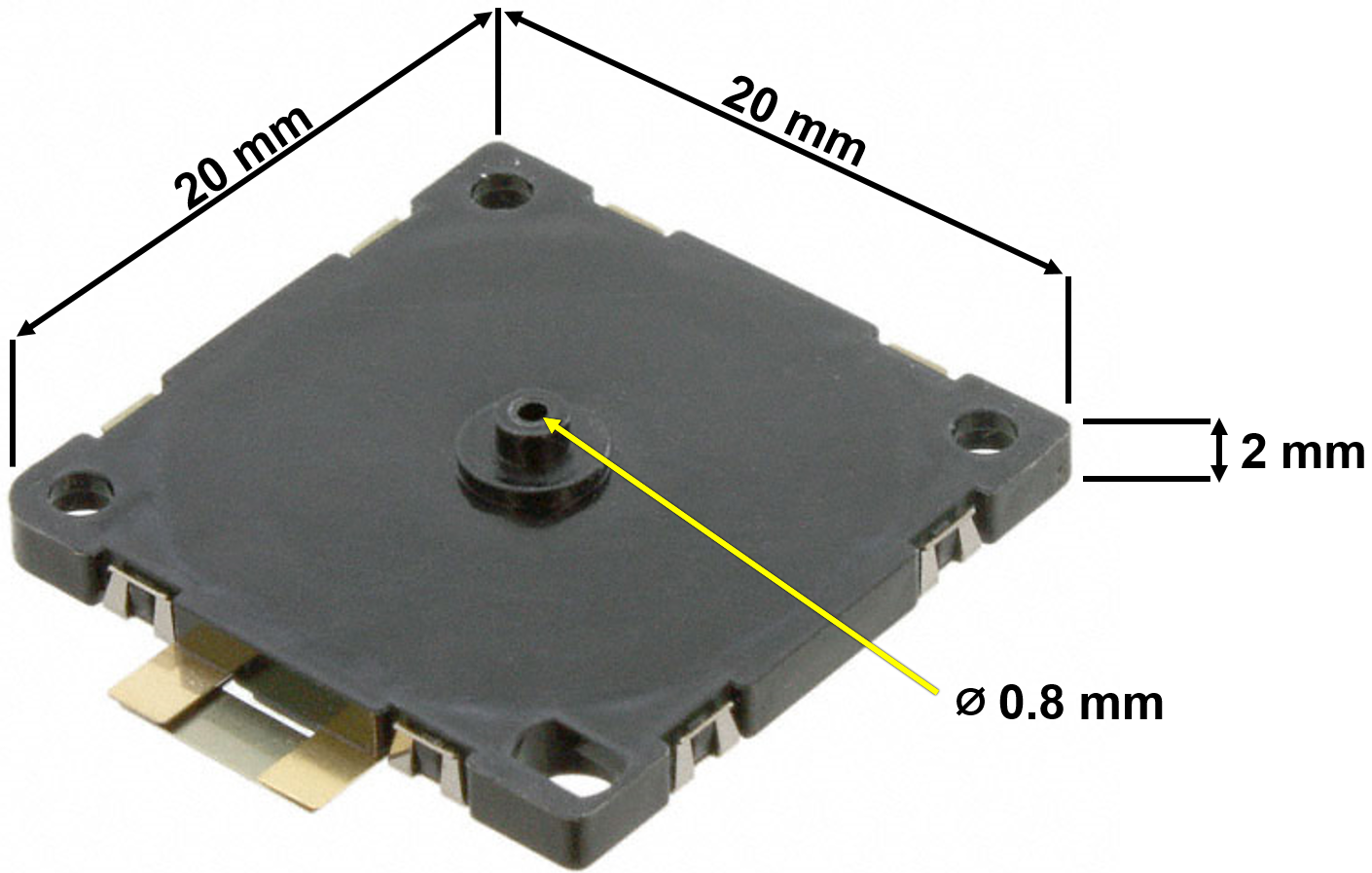}
    \caption{Murata MZB1001T02 microblower
    \label{fig:SJA}}
\end{figure}

\begin{figure}
    \centering
    \includegraphics[width=0.7\linewidth]{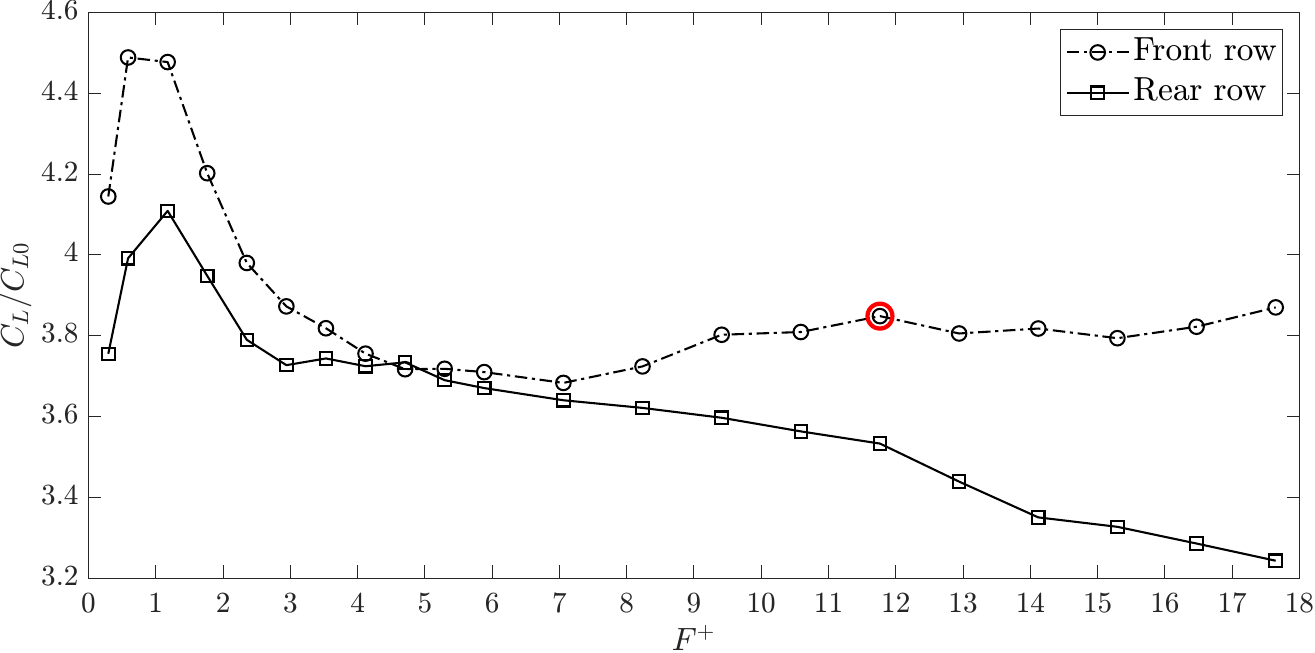}
    \caption{Normalized lift coefficient versus reduced frequency for front and rear row actuation. Adapted from \citet{Xu2025}\label{fig:liftVsfrequency}}
\end{figure}

The SJAs used are the commercially available Murata MZB1001T02 microblowers, pictured in Figure~\ref{fig:SJA}, and are flush-mounted beneath the surface of the airfoil. The SJA nozzles are equally spaced in the spanwise direction by \SI{25}{\milli\metre}. The actuator spacing was minimized based on the physical size of the MZB1001T02 microblowers, with a small allowance for wiring and integration. The array consists of two rows of 12 SJAs located at \SI{10.7}{\percent} and \SI{19.8}{\percent} chord, positioned to place one row upstream and one downstream of the separation point under the flow conditions of interest. Prior studies have shown that actuation upstream of separation is more effective for flow control~\cite{Feero2017b,Xu2025}; therefore, only the upstream row was activated in these experiments, as indicated by the arrows in Figure~\ref{fig:airfoil}. Figure~\ref{fig:liftVsfrequency} presents previously documented lift coefficients for both the upstream and downstream rows across a range of tested modulation frequencies, highlighting the greater effectiveness of the front row. The SJA has a specified operating range of 5--30~$\mathrm{V_{pp}}$ and a drive resonant frequency between 24--27~\unit{\kilo\hertz}. The mean centerline velocity of the synthetic jet reached a maximum when driven at a frequency of \SI{25.1}{\kilo\hertz}. However, due to the right-skewed velocity response, the carrier frequency was chosen as $f_c=25.5$~\unit{\kilo\hertz} to ensure a stable jet velocity~\cite{Xu2023}.

The SJAs were burst modulated at an excitation frequency of $f_m=200$~\unit{\hertz}, corresponding to a non-dimensional frequency of $F^+=11.76$, which targets the Kelvin-Helmholtz instability in the separated shear layer~\cite{Feero2017b,Yarusevych2009a}. Prior parametric studies showed that forcing at this frequency results in a local maximum in lift coefficient, indicating strong flow receptivity to these control parameters, as highlighted in Figure~\ref{fig:liftVsfrequency}~\cite{Xu2025}. While forcing at $F^+\approx\mathcal{O}(1)$ yields higher mean lift, it produces significant unsteadiness~\cite{Machado2024b}. High-frequency actuation ($F^+=11.76$) was therefore selected for its stable response and favorable drag characteristics~\cite{Xu2023}, enabling the present study to systematically examine the effect of duty cycle on both lift and flow stability. Square waveforms were used for the carrier and modulation signals. The DC was varied from \SI{5}{\percent} to \SI{95}{\percent} and the blowing ratio was varied from $C_B=1.9$ to $C_B=5.0$. The time-averaged jet velocities used to calculate the blowing ratio were measured previously by \citet{Chovet2016}. These blowing ratios were achieved by varying the input voltage from 10--20~$\mathrm{V_{pp}}$. Previous visualizations showed that at 5~$\mathrm{V_{pp}}$, the flow controlled by the SJA remains highly asymmetric even at high duty cycles, due to increased sensitivity in the actuator’s velocity response at low voltage amplitudes~\cite{Ho2024}. Consequently, this study focuses on operating voltages of 10~$\mathrm{V_{pp}}$ and above to ensure the analysis captures fluidic effects while minimizing random variations in the SJAs' response. The fixed control parameters of the SJA are summarized in Table~\ref{tab:control_parameters}, while Table~\ref{tab:exp_matrix} lists the experimental matrix of varied control parameters and the resulting momentum coefficients.

\begin{table}
\caption{Summary of constant control parameters}
\label{tab:control_parameters}
\centering
\begin{tabular}{ll}
\hline\hline
        SJA nozzle diameter & $d=\SI{0.8}{\milli\metre}$ \\
        Chordwise position of activated SJAs & $x_{\mathrm{jet}}=0.107c$ \\
        Number of SJAs & 12 \\
        Spanwise SJA spacing & \SI{25}{\milli\metre} \\
        Waveform & Square \\
        Carrier frequency & $f_c=\SI{25.5}{\kilo\hertz}$ \\
        Modulation frequency & $f_m=\SI{200}{\hertz}$ ($F^+=11.76$) \\
\hline\hline
\end{tabular}
\end{table}

\begin{table}
\centering
\caption{Experimental matrix of tested control cases and corresponding momentum coefficients}
\label{tab:exp_matrix}
\begin{threeparttable}
\centering
\renewcommand{\arraystretch}{1.2}
\begin{tabular}{cccc}
\hline\hline
Voltage (\si{\volt}) & $C_B$ & $DC$ (\si{\percent}) & $C_\mu$ \\ 
\hline
\multirow{4}{*}{10} & \multirow{4}{*}{1.9} & 5.0  & $1.5\times10^{-6}$\tnote{¶}\\
                    &                       & 12.5 & $3.7\times10^{-6}$ \\
                    &                       & 37.5 & $2.4\times10^{-4}$ \\
                    &                       & 62.5 & $5.0\times10^{-4}$ \\
                    &                       & 95   & $7.9\times10^{-4}$ \\
\hline
\multirow{4}{*}{15} & \multirow{4}{*}{4.4} & 5.0  & $2.0\times10^{-5}$\tnote{¶}\\
                    &                       & 12.5 & $5.0\times10^{-5}$ \\
                    &                       & 37.5 & $6.1\times10^{-4}$ \\
                    &                       & 62.5 & $1.1\times10^{-3}$ \\
                    &                       & 95   & $1.5\times10^{-3}$ \\
\hline
\multirow{5}{*}{20} & \multirow{5}{*}{5.0} & 5.0  & $3.0\times10^{-6}$ \\
                    &                       & 12.5 & $1.5\times10^{-4}$ \\
                    &                       & 37.5 & $9.0\times10^{-4}$ \\
                    &                       & 62.5 & $1.5\times10^{-3}$ \\
                    &                       & 95   & $2.0\times10^{-3}$ \\
\hline\hline
\end{tabular}
\begin{tablenotes}
    \item[¶]Scaled linearly from \SI{12.5}{\percent} DC
\end{tablenotes}
\end{threeparttable}
\end{table}

\begin{figure}
    \centering
    \includegraphics[width=0.6\linewidth]{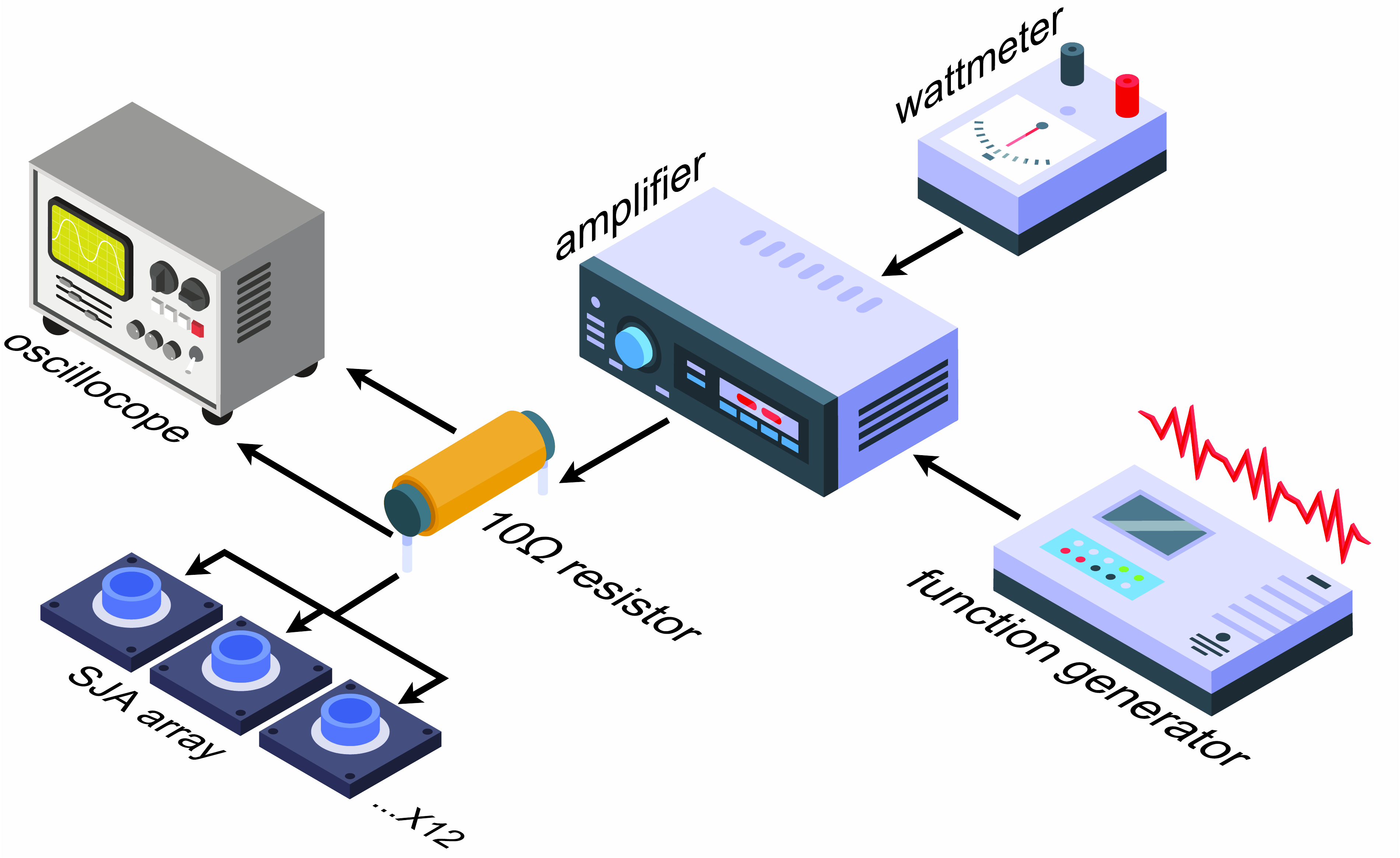}
    \caption{Input signal configuration and power measurement setup for the SJA array}
    \label{fig:electrical_diagram}
\end{figure}

Figure~\ref{fig:electrical_diagram} illustrates the setup in which the SJA array receives its input signal from a Rigol DG1022Z function generator, amplified by a YAMAHA HTR5470 amplifier. The signal is then routed through a 10~$\Omega$ resistor to prevent overloading before reaching the 12 SJAs wired in parallel. The power consumption of the SJA array was determined by measuring the power consumed by the amplifier while the SJAs were active ($P_\mathrm{amp,active}$) and subtracting the idle power consumption of the amplifier ($P_\mathrm{amp,idle}$) and the power dissipated by the resistor ($P_\mathrm{resistor}$):
\begin{equation}
    P_\mathrm{SJA}=P_\mathrm{amp,active}-P_\mathrm{amp,idle}-P_\mathrm{resistor}.
\end{equation}
The power consumption of the amplifier was measured using a consumer-grade wattmeter, which has an uncertainty of $\pm$\SI{0.1}{\watt}. Power readings were monitored for at least one minute to ensure stability within the instrument's stated uncertainty. A Rigol DS1202Z-E oscilloscope was used to measure the RMS voltage drop across the resistor. The power dissipated by the resistor was then calculated using the equation:
\begin{equation}
    P_\mathrm{resistor}=\frac{V_\mathrm{RMS}^2}{R},
\end{equation}
where $V_\mathrm{RMS}$ is the measured RMS voltage drop across the resistor, and $R$ is the resistance value. The uncertainty in this calculation was determined to be $\pm$\SI{0.1}{\watt}. Therefore, the total uncertainty in the power consumption of the SJAs was estimated to be $\pm$\SI{0.2}{\watt}.

Hot-wire anemometry (HWA) was used to characterize the microblower’s response for each control case and to calculate the corresponding momentum coefficient. The probe was positioned \SI{4}{\milli\meter} ($5d$) downstream of the orifice and sampled at \SI{10}{\kilo\hertz} for \SI{20}{\second} at each point. At the lowest duty cycle (\SI{5}{\percent}), the jet velocity fell below the detection threshold of the hot-wire, except at the highest input voltage (20~$\mathrm{V_{pp}}$).

\begin{figure}
    \begin{subfigure}{0.44\linewidth}
    \centering
    \includegraphics[width=\linewidth]{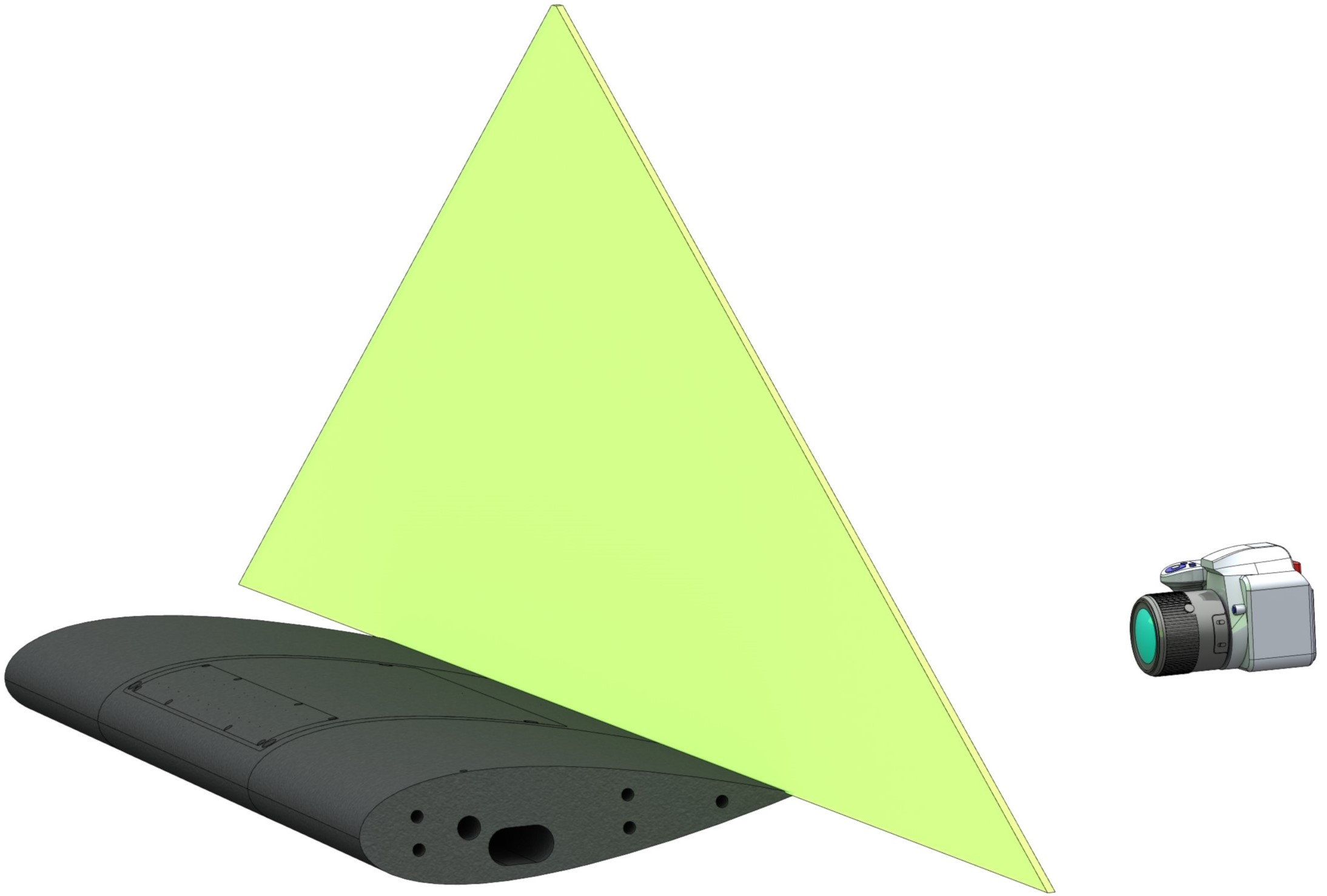}
    \caption{Laser sheet and camera orientation for smoke flow visualization}
    \label{fig:laser_sheet_setup}
    \end{subfigure}
    \hfill
    \begin{subfigure}{0.55\linewidth}
    \centering
    \includegraphics[width=\linewidth]{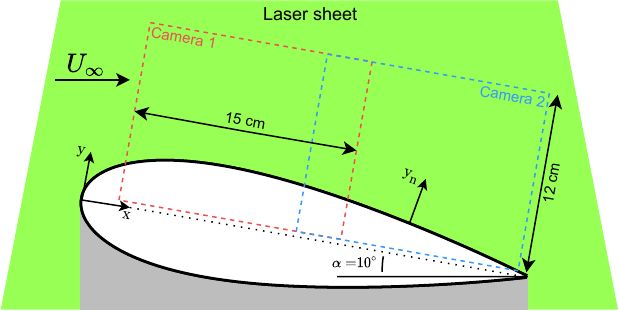}
    \caption{PIV camera FOVs and laser sheet orientation}
    \label{fig:pivsetup}
    \end{subfigure}
    \caption{Experimental setup for visualization and measurement}
\end{figure}

Cross-sectional smoke flow visualization in the spanwise-transverse plane was performed with a single horizontal smoke wire upstream of the wing. The horizontal smoke wire was positioned just above the airfoil’s stagnation point to visualize the flow at the edge of the shear layer. The horizontal smoke wire was installed \SI{9.5}{\centi\metre} upstream of the leading edge, as shown in blue in Figure~\ref{fig:airfoil}. The smoke wire length spanned slightly larger than the SJA array. Figure~\ref{fig:airfoil} shows how the horizontal wire (blue dashed line) is installed between two vertical support wires (yellow dash-dot lines). The horizontal wire was kept taut by making it slightly shorter than the distance between the vertical support wires so that they bowed inward. This also helped ensure the horizontal wire remained in the same position across all tests. The wire was coated with oil and then heated resistively to generate smoke streaks. A laser sheet, oriented perpendicular to the flow, illuminated the generated smoke at the trailing edge, providing a sectional visualization of the shear layer boundary (Figure~\ref{fig:laser_sheet_setup}). A Nikon D7000 DSLR camera downstream of the airfoil and test section was used to image the smoke streaks. The camera was operated in burst mode, capturing up to six images per second, facilitating the capture of the smoke visualization at peak density. A long exposure time of 1/6~\unit{\second} (2.8 convective timescales) was used to provide a sense of the mean flow while capturing unsteady areas highlighted by motion blur. Additionally, for each control case, 3–6 images were captured in which the attached region was clearly defined. Minimal variability was observed across these images; the shape and extent of the attached region remained consistent, with differences smaller than the thickness of the illuminated smoke streaks. The camera was operated with an aperture of $f/$1.8, and an ISO of 4000. Additional details on the smoke visualization method can be found in \citet{Machado2024b,Machado2024a}.

Particle image velocimetry (PIV) measured the streamwise-transverse velocity field above the airfoil. Neutrally buoyant particles were introduced into the flow using a fog machine downstream of the test section. Two JAI SP-5000M-USB cameras, each with a resolution of 2560~$\times$~2048, were positioned outside the test section and aligned with the airfoil's chord. Composite images were created by stitching together the camera captures, resulting in a field of view (FOV) measuring \SI{270}{\milli\metre}~$\times$~\SI{120}{\milli\metre}, with a pixel resolution of \SI{17}{pixels\per\milli\metre} (Figure~\ref{fig:pivsetup}). The stitched FOV effectively captured the range $x/c \in [0.1,1]$ and $y/c \in [0,0.4]$, covering the area of interest above the airfoil. A Litron Bernoulli \SI{200}{\milli\joule} Nd-YAG laser with a wavelength of \SI{532}{\nano\metre} was passed through converging and diverging cylindrical lenses ($f=1000$~\unit{\milli\metre} and $f=-13.7$~\unit{\milli\metre}, respectively) to create a thin laser sheet that illuminated the measurement plane. The image acquisition and laser pulses were synchronized using an NI PCIe-6323 data acquisition card (DAQ) at \SI{10}{\hertz}, and 1000 image pairs were recorded for each measurement. The time delay between the frames in an image pair was set to $\SI{120}{\micro\second}$. The uncertainty in the mean velocity was calculated using the standard error, $1.96\sigma/\sqrt{N}$, where $\sigma$ is the local standard deviation and $N$ is the number of samples, corresponding to a \SI{95}{\percent} confidence interval. For the baseline separated flow case, which represents the highest expected variability due to turbulence, this uncertainty remained below $0.03U_\infty$ throughout the field. For the control cases, phase-locked velocity measurements were achieved by synchronizing image acquisition at eight evenly spaced phases relative to the SJA modulation frequency, where a phase angle of $\phi=0$\si{\degree} represents the activation of the actuator. Coherent fluctuations are extracted from the velocity signal using the triple decomposition~\cite{Hussain1970}. The streamwise velocity was decomposed as $u=\bar{u}+\tilde{u}+u'$, where $\bar{u}$ is the time-averaged velocity, $\tilde{u}$ is the coherent velocity obtained from the phase-average, and $u'$ is the fluctuation velocity~\cite{Buchmann2013}. Velocity vectors were obtained using open-source software OpenPIV-Python-GPU, utilizing a multi-pass cross-correlation algorithm to accurately record large and small displacements. The PIV process involved an initial iteration at an interrogation window size of 64~$\times$~64 pixels, followed by two iterations at 32~$\times$~32 pixels, and two iterations at 16~$\times$~16 pixels. Linear window deformation was used to reduce correlation errors in high-shear regions.

\section{Results and Discussion}
\subsection{Microblower characterization}
\label{characterization}
\begin{figure}
\centering
    \includegraphics[width=0.5\linewidth]{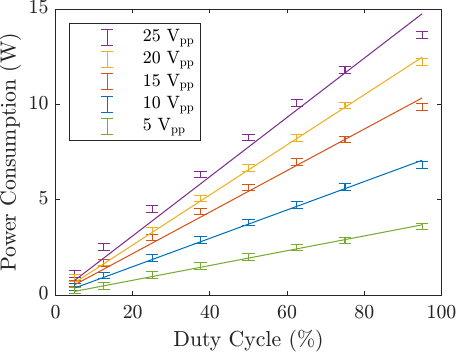}
    \caption{Power consumption of the SJA array with the front-row actuators active}
    \label{fig:power_consumption}
\end{figure}

Figure~\ref{fig:power_consumption} presents the power consumption of the SJA array, measured at various operating voltages and DCs. The results reveal a clear linear dependence of the power consumption on the DC, as highlighted by the lines of best fit.  Secondly, an increase in voltage leads to higher power consumption and a steeper slope in the relationship between power consumption and DC, i.e. $\frac{\mathrm{d}P}{\mathrm{d}(DC)}$ increases, as expected.

\begin{figure}
\centering
\begin{subfigure}{0.495\linewidth}
        \includegraphics[width=\linewidth]{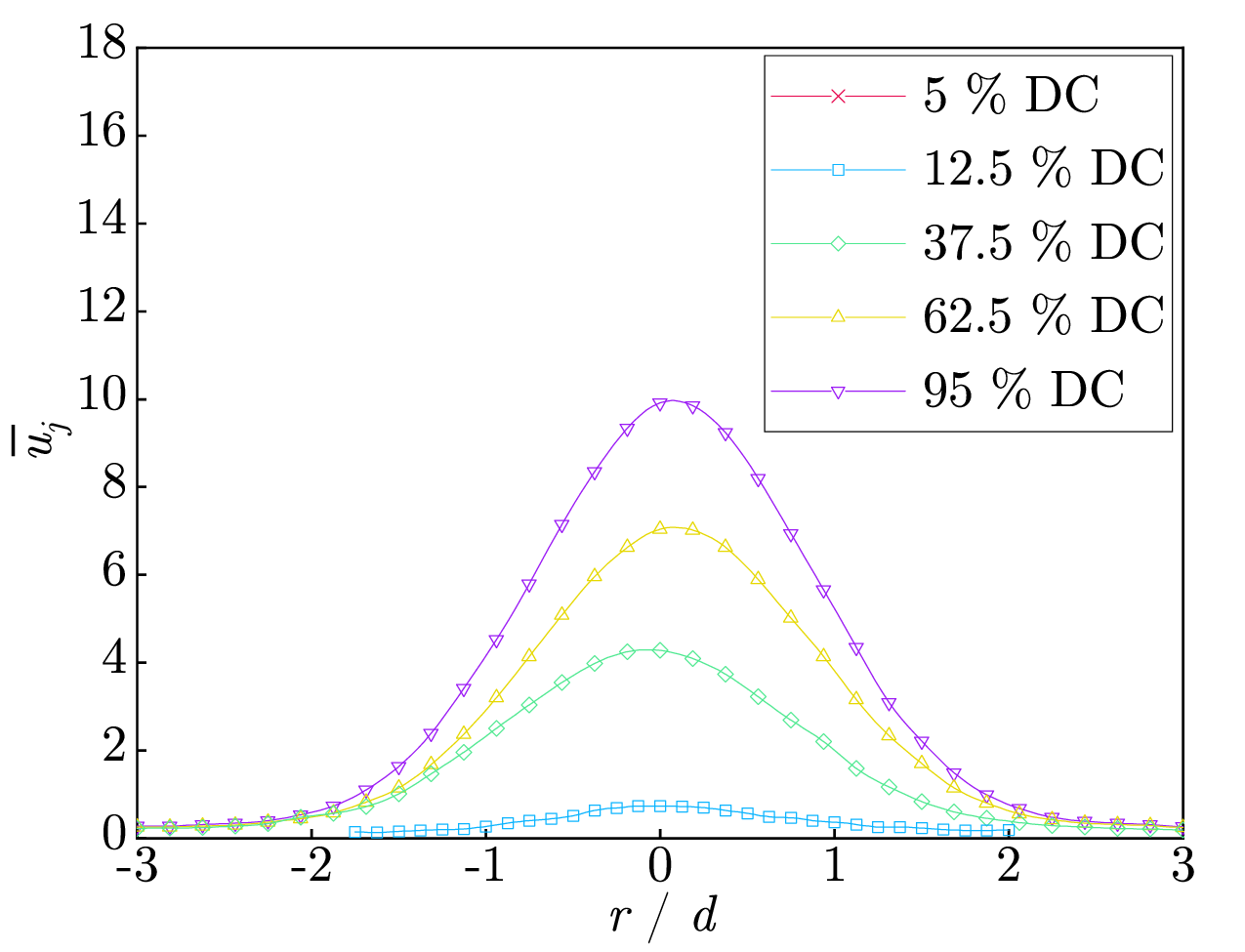}
        \caption{$\mathrm{V_{pp}}=10$}
        \label{fig:U_10V}
    \end{subfigure}
    \begin{subfigure}{0.495\linewidth}
        \includegraphics[width=\linewidth]{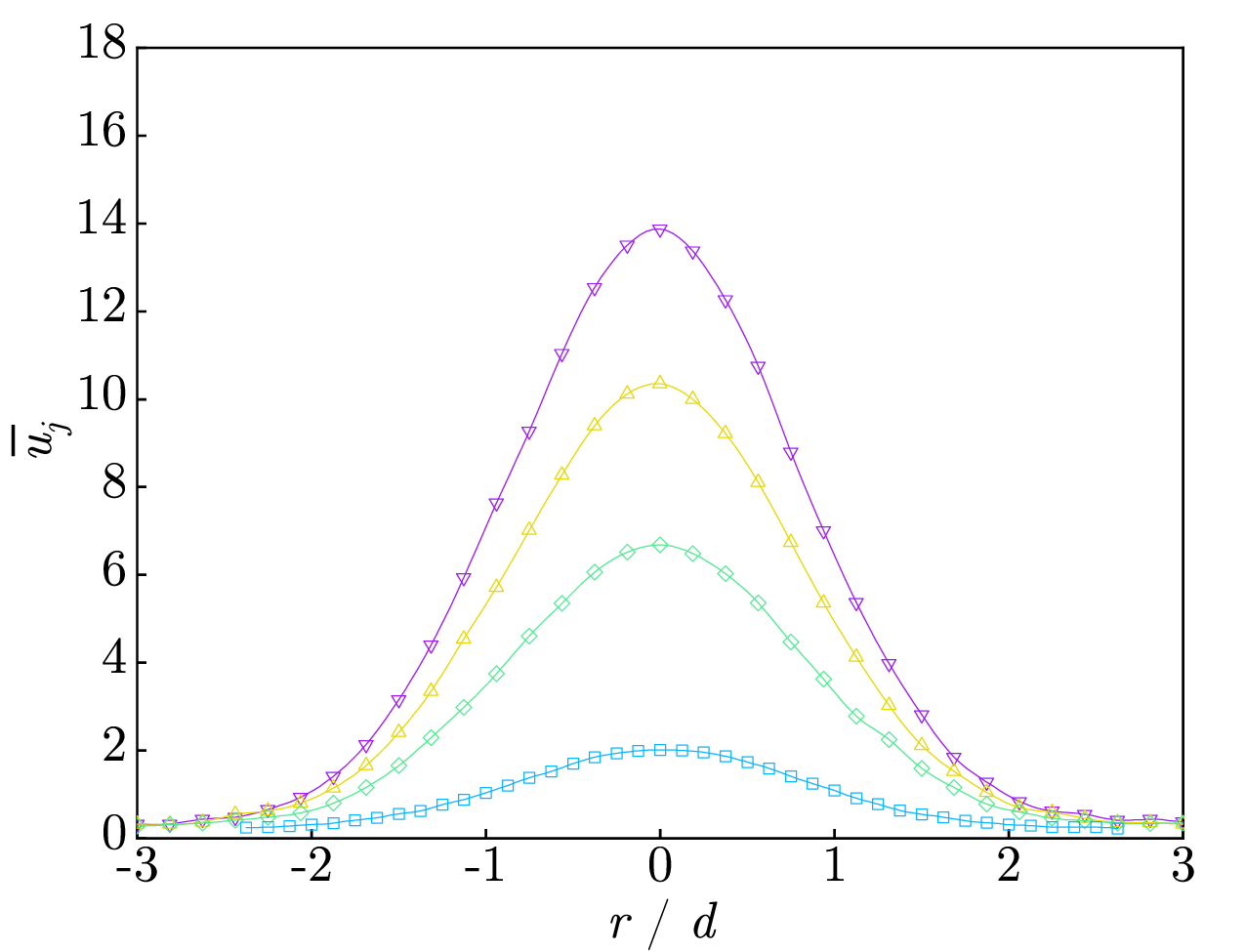}
        \caption{$\mathrm{V_{pp}}=15$}
        \label{fig:U_15V}
    \end{subfigure}
    \begin{subfigure}{0.495\linewidth}
        \includegraphics[width=\linewidth]{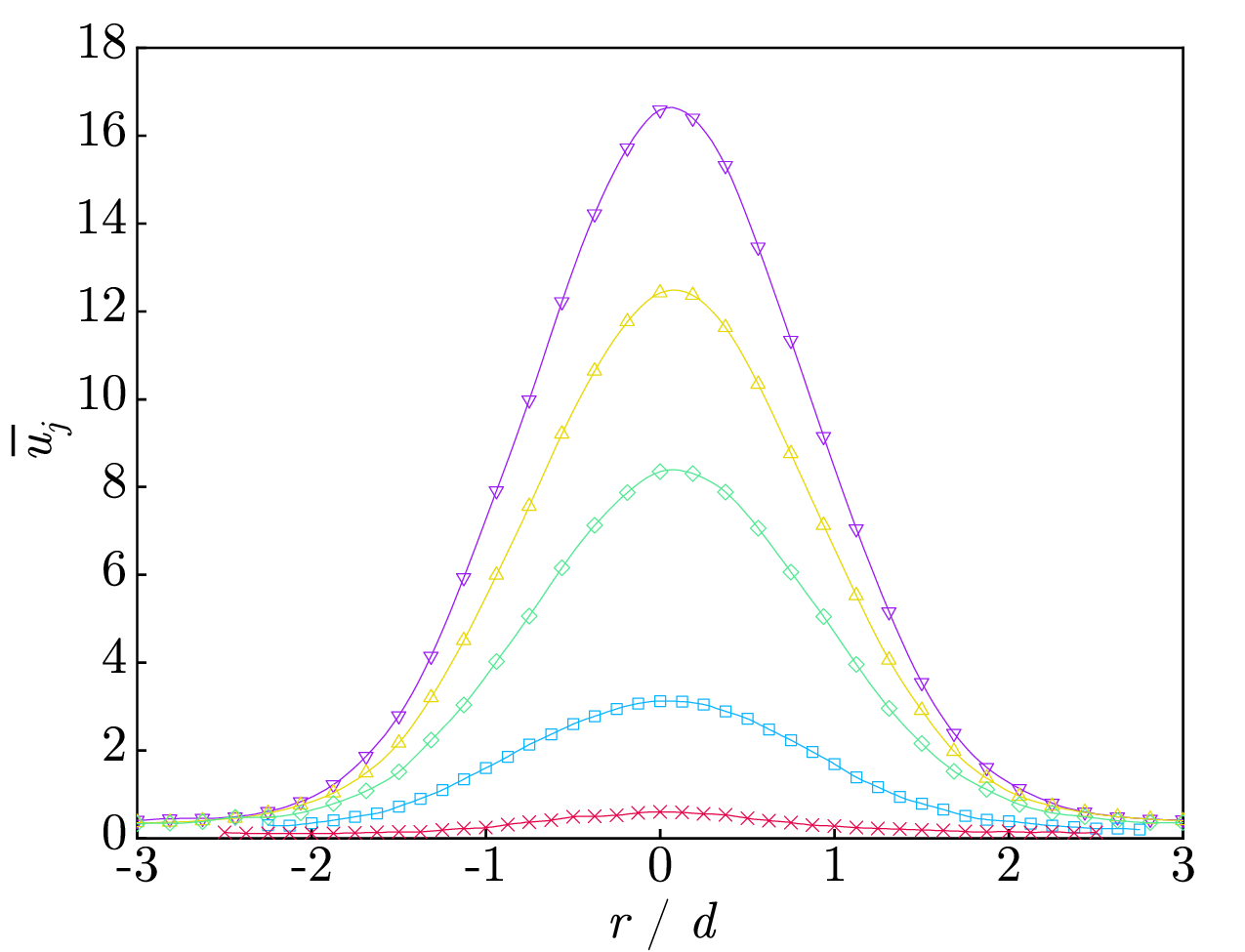}
        \caption{$\mathrm{V_{pp}}=20$}
        \label{fig:U_20V}
    \end{subfigure}
    \caption{Time-averaged jet velocity profiles measured at $5d$ from the exit by HWA}
    \label{fig:U_DC}
\end{figure}

\begin{figure}
\centering
\begin{subfigure}{0.495\linewidth}
        \includegraphics[width=\linewidth]{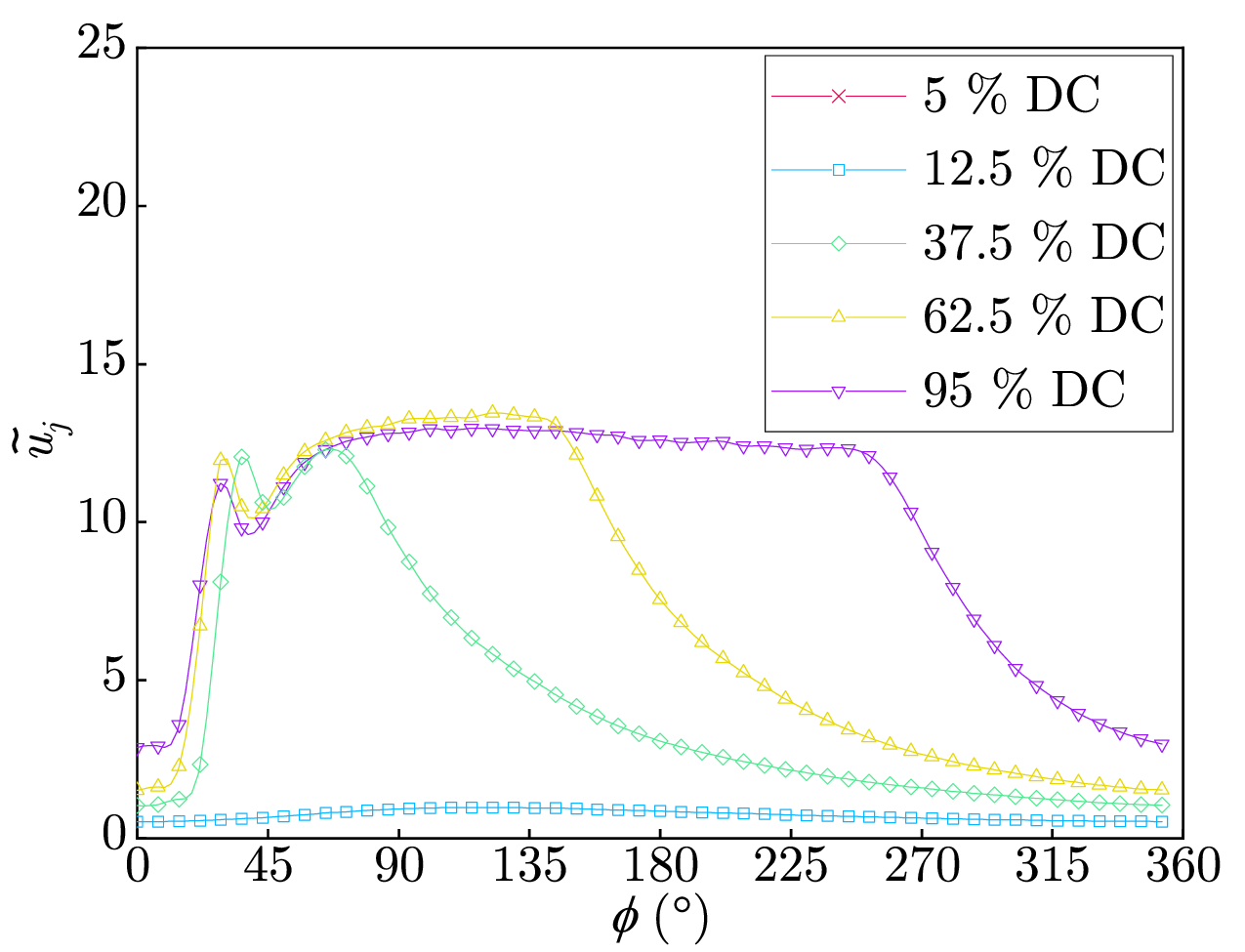}
        \caption{$\mathrm{V_{pp}}=10$}
        \label{fig:Up_10V}
    \end{subfigure}
    \begin{subfigure}{0.495\linewidth}
        \includegraphics[width=\linewidth]{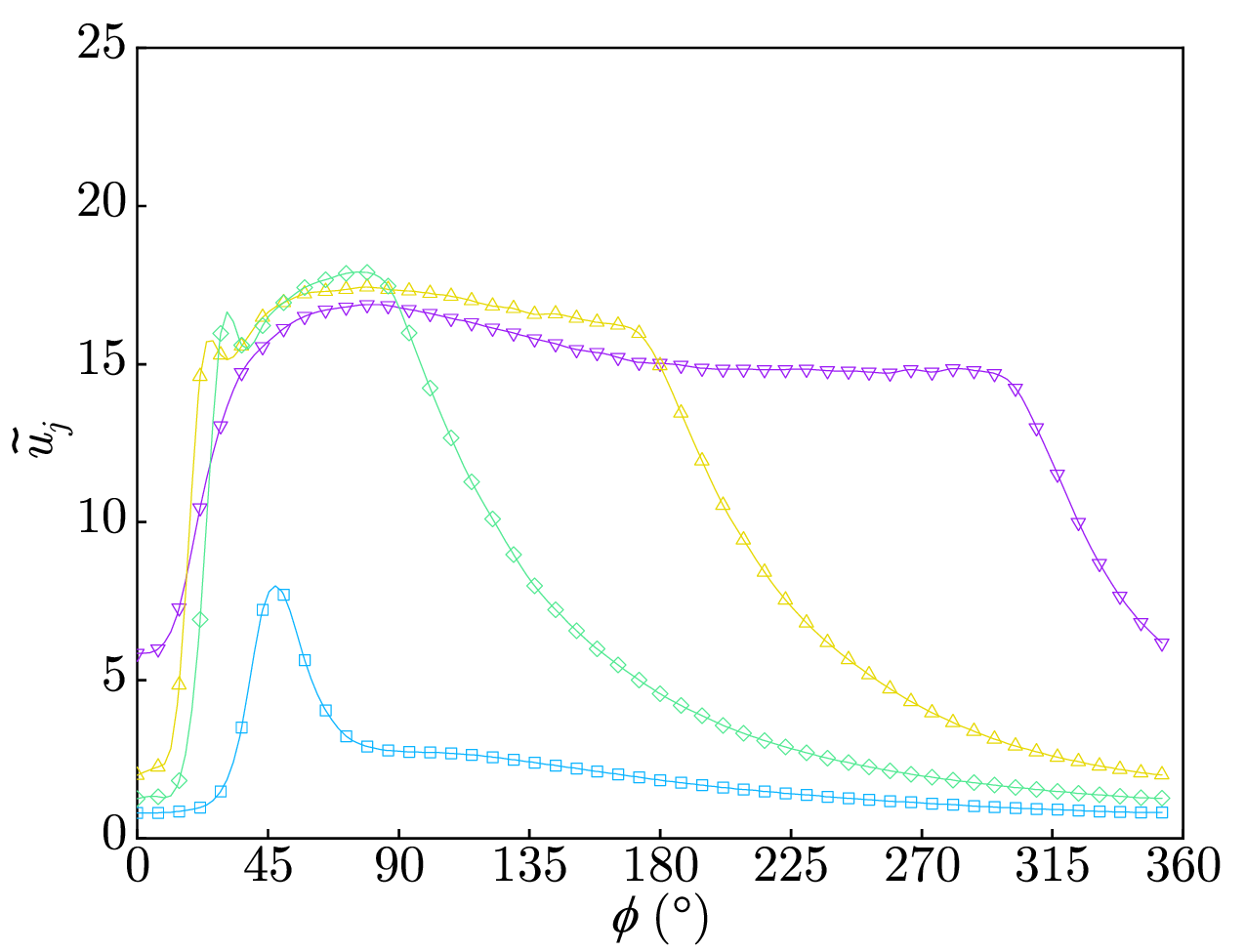}
        \caption{$\mathrm{V_{pp}}=15$}
        \label{fig:Up_15V}
    \end{subfigure}
    \begin{subfigure}{0.495\linewidth}
        \includegraphics[width=\linewidth]{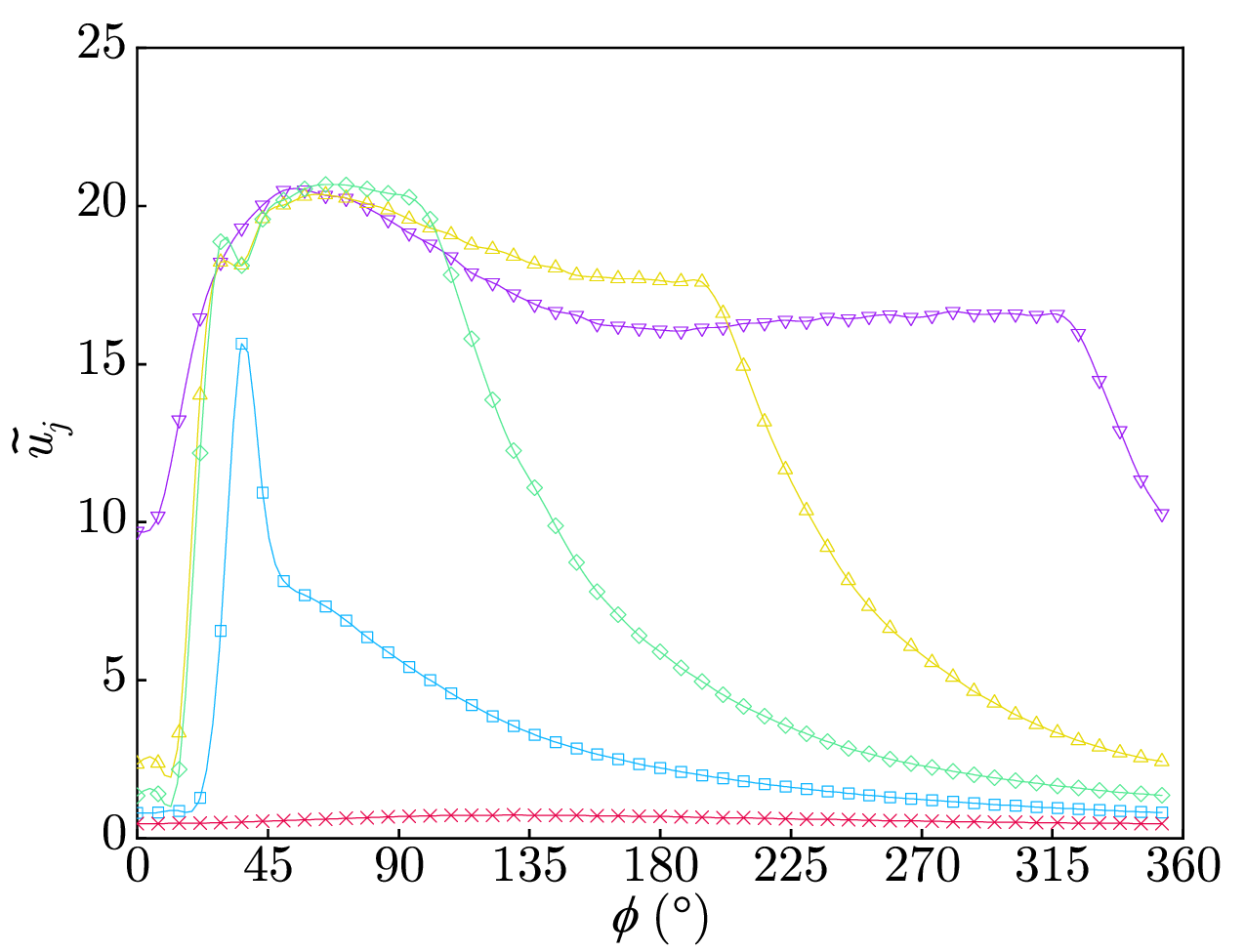}
        \caption{$\mathrm{V_{pp}}=20$}
        \label{fig:Up_20V}
    \end{subfigure}
    \caption{Phase-averaged centerline velocity of the jet at $f_c = \SI{25.5}{\kilo\hertz}$ and $f_m = \SI{200}{\hertz}$, measured $5d$ from the exit by HWA}
    \label{fig:U_phase}
\end{figure}

\begin{figure}
    \centering
    \includegraphics[width=0.5\linewidth]{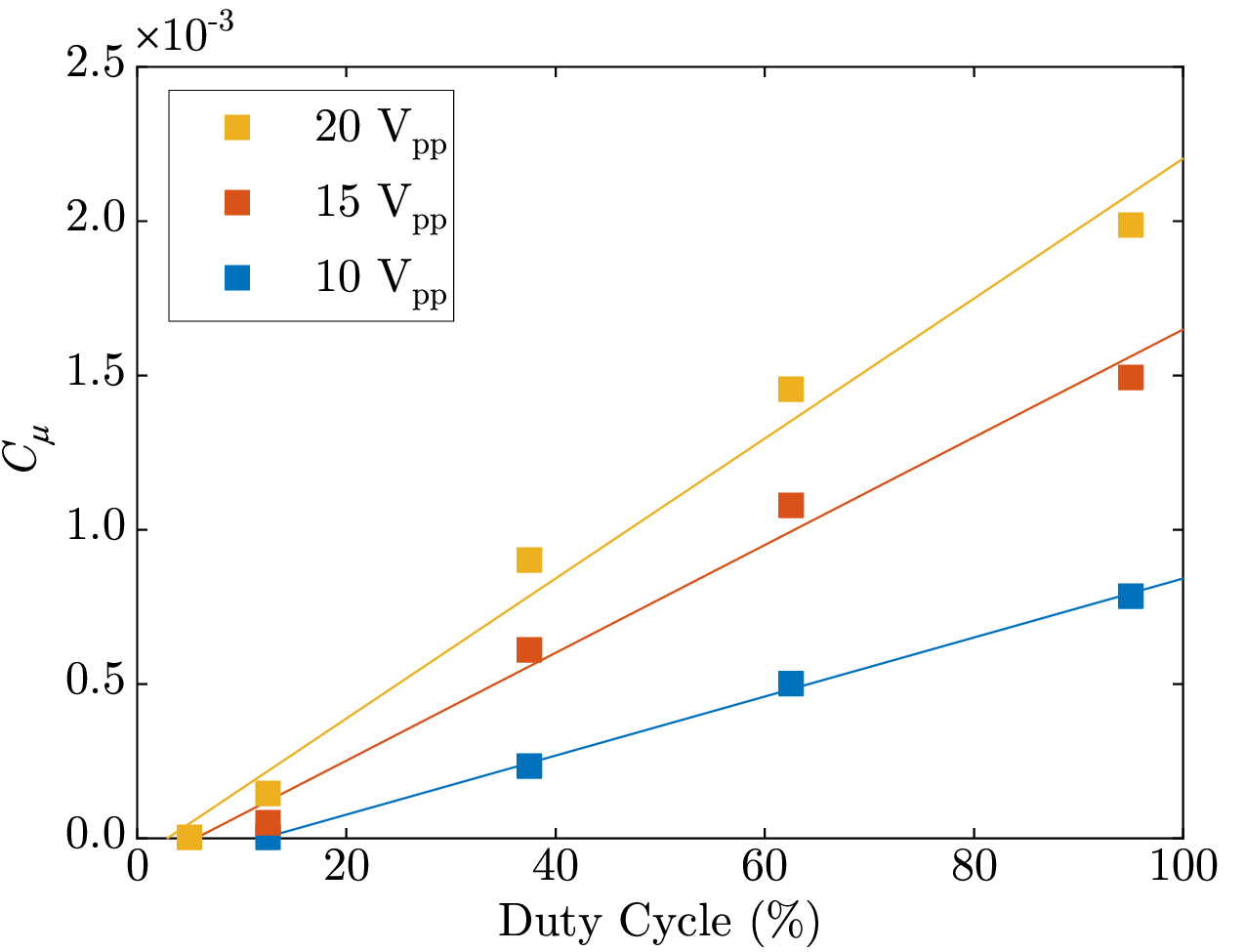}
    \caption{Momentum coefficient at various input voltages and DCs}
    \label{fig:Cu}
\end{figure}

Figure~\ref{fig:U_DC} presents time-averaged velocity profiles of the synthetic jet against radial distance, measured by HWA $5d$ downstream of the microblower orifice. The temporal evolution of the jet is presented in Figure~\ref{fig:U_phase}, where the phase-averaged centerline velocity is plotted over a complete burst cycle. It is evident that the microblower response to the input square wave does not follow a square waveform, and the phase-averaged velocity remains nonzero during the inactive portion of the burst cycle. This indicates a finite decay time of the jet rather than an instantaneous shutoff. Despite this transient response, the momentum coefficient exhibits a linear scaling with the duty cycle, as demonstrated in Figure~\ref{fig:Cu}. A similar linear relationship between duty cycle and time-averaged momentum has been observed in plasma-based actuators at constant frequency \cite{Hui2022}.

\subsection{Aerodynamic Response to Duty Cycle and Blowing Ratio}
\label{sec:parametric}
\begin{figure}
\centering
\begin{subfigure}{0.495\linewidth}
        \includegraphics[width=\linewidth]{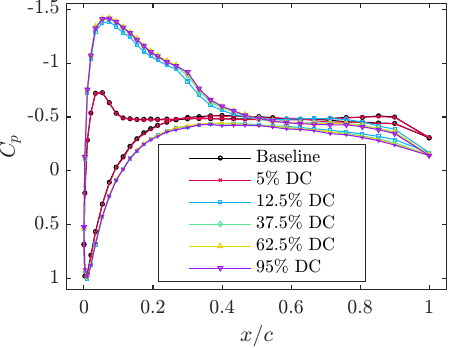}
        \caption{$C_B=1.9$}
        \label{fig:pressure_10V}
    \end{subfigure}
    \begin{subfigure}{0.495\linewidth}
        \includegraphics[width=\linewidth]{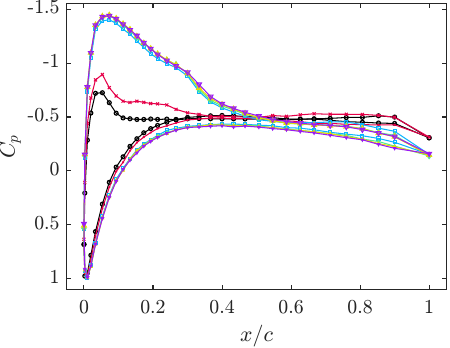}
        \caption{$C_B=4.4$}
        \label{fig:pressure_15V}
    \end{subfigure}
    \begin{subfigure}{0.495\linewidth}
        \includegraphics[width=\linewidth]{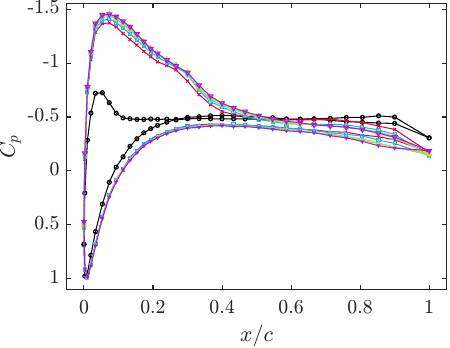}
        \caption{$C_B=5.0$}
        \label{fig:pressure_20V}
    \end{subfigure}
    \caption{Surface pressure distributions for various control strategies, compared with the baseline case}
    \label{fig:pressure_distributions}
\end{figure}

Surface pressure distributions along the midspan are plotted in Figure~\ref{fig:pressure_distributions} for the baseline case and the various control strategies. Stall is evident in the baseline case, characterized by a plateau in the $C_p$ distribution of the suction surface, extending from slightly downstream of the suction peak to the trailing edge. At $C_B=1.9$ (Figure~\ref{fig:pressure_10V}), all tested DCs except for \SI{5}{\percent} result in a fully reattached flow, as indicated by the absence of a pressure plateau in the $C_p$ distribution. Additionally, lift recovery is evident from the heightened suction peaks. The \SI{5}{\percent} DC shows a $C_p$ distribution nearly identical to the baseline case, suggesting that this control strategy has a negligible effect on the lift. Increasing the blowing ratio to $C_B=4.4$ improves the pressure distribution for the \SI{5}{\percent} DC case slightly; however, the flow is still separated, as indicated by the pressure plateau downstream of $x/c=0.3$ (Figure~\ref{fig:pressure_15V}). Further increasing the blowing ratio to $C_B=5.0$ results in complete flow reattachment for all DCs, including \SI{5}{\percent}. This demonstrates that the threshold DC for flow reattachment is dependent on the blowing ratio, consistent with literature indicating that a minimum momentum coefficient is required to fully reattach the flow~\cite{Goodfellow2013,Feero2017b}. A review by \citet{Greenblatt2022} states that once a threshold momentum coefficient is attained, significant lift increases are observed. In the present study, this momentum threshold is surpassed by increasing either the blowing ratio or DC. For DCs above \SI{12.5}{\percent}, flow reattachment is observed at all blowing ratios. Once reattachment is achieved, further increases in the DC or blowing ratio provide only marginal additional lift enhancement, as indicated by the slightly increased suction pressures.

\begin{figure}
    \centering
    \begin{subfigure}{\linewidth}
        \centering
        \includegraphics[width=0.302\linewidth]{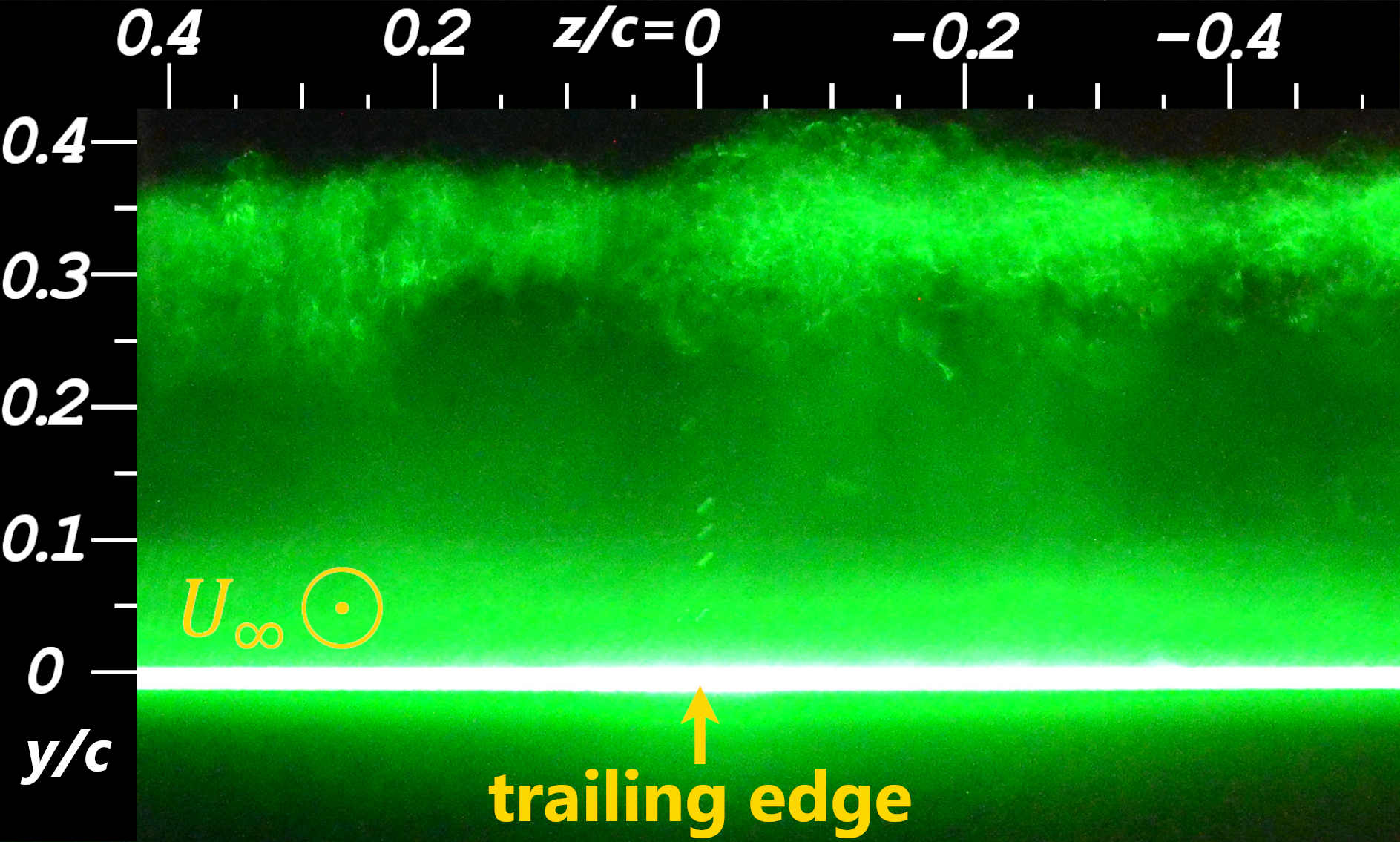}
        \caption{Baseline}
        \label{fig:smoke_baseline}
    \end{subfigure}
    \newline
    \begin{subfigure}{0.302\linewidth}
        \includegraphics[width=\linewidth]{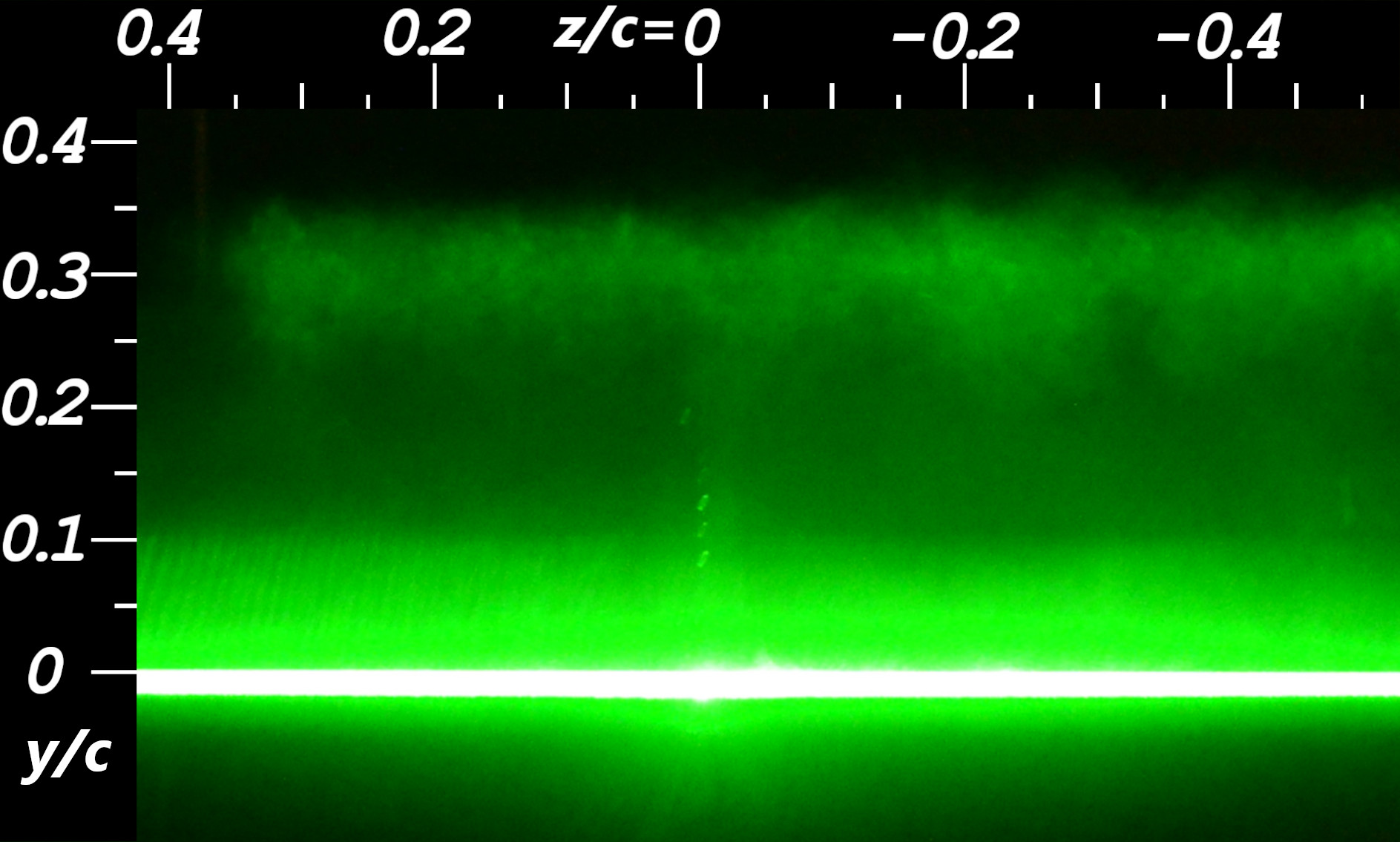}
        \caption{$C_B=1.9$; \SI{5}{\percent} DC}
        \label{fig:smoke_10V_5DC}
    \end{subfigure}
    \begin{subfigure}{0.302\linewidth}
        \includegraphics[width=\linewidth]{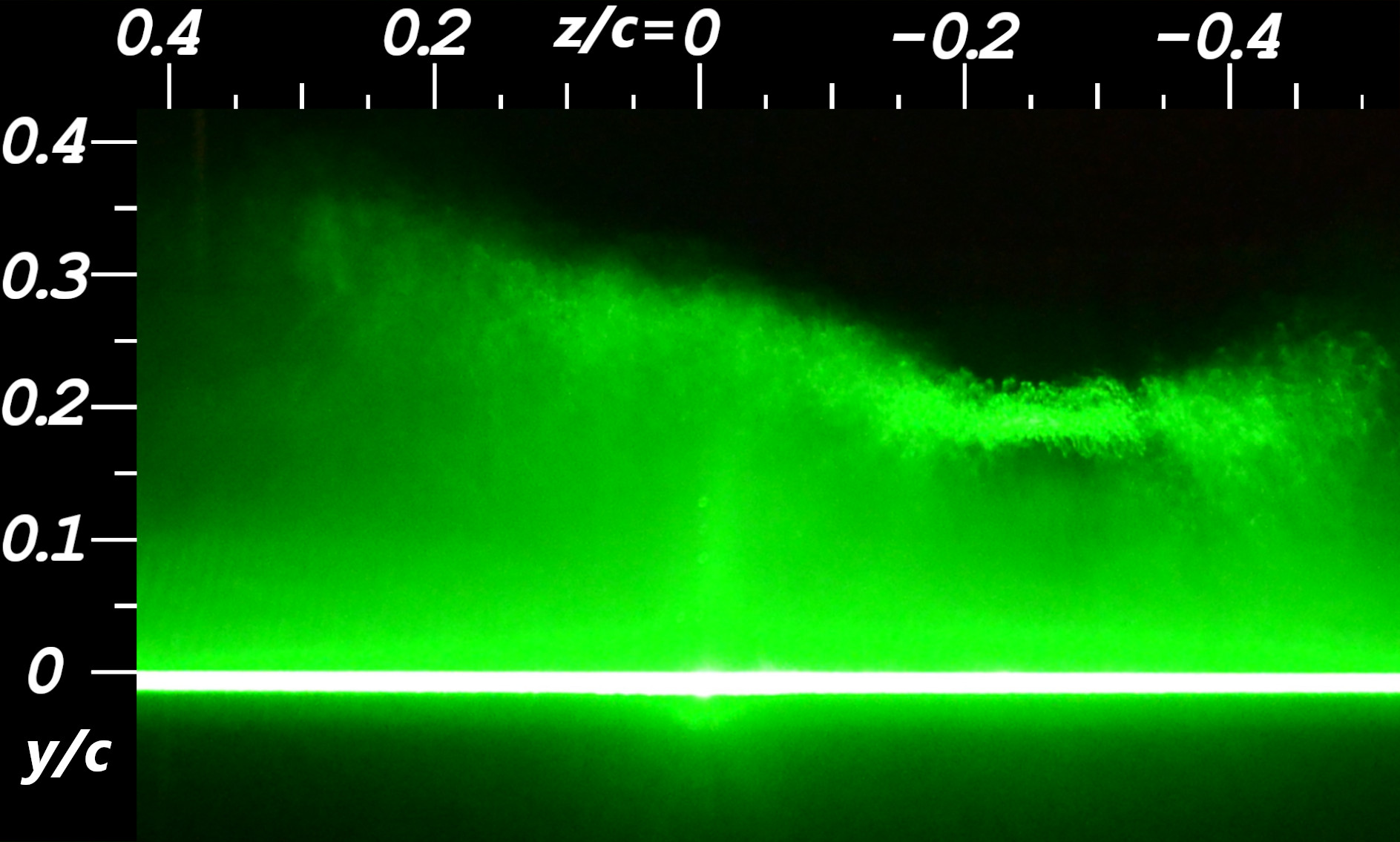}
        \caption{$C_B=4.4$; \SI{5}{\percent} DC}
        \label{fig:smoke_15V_5DC}
    \end{subfigure}
    \begin{subfigure}{0.302\linewidth}
        \includegraphics[width=\linewidth]{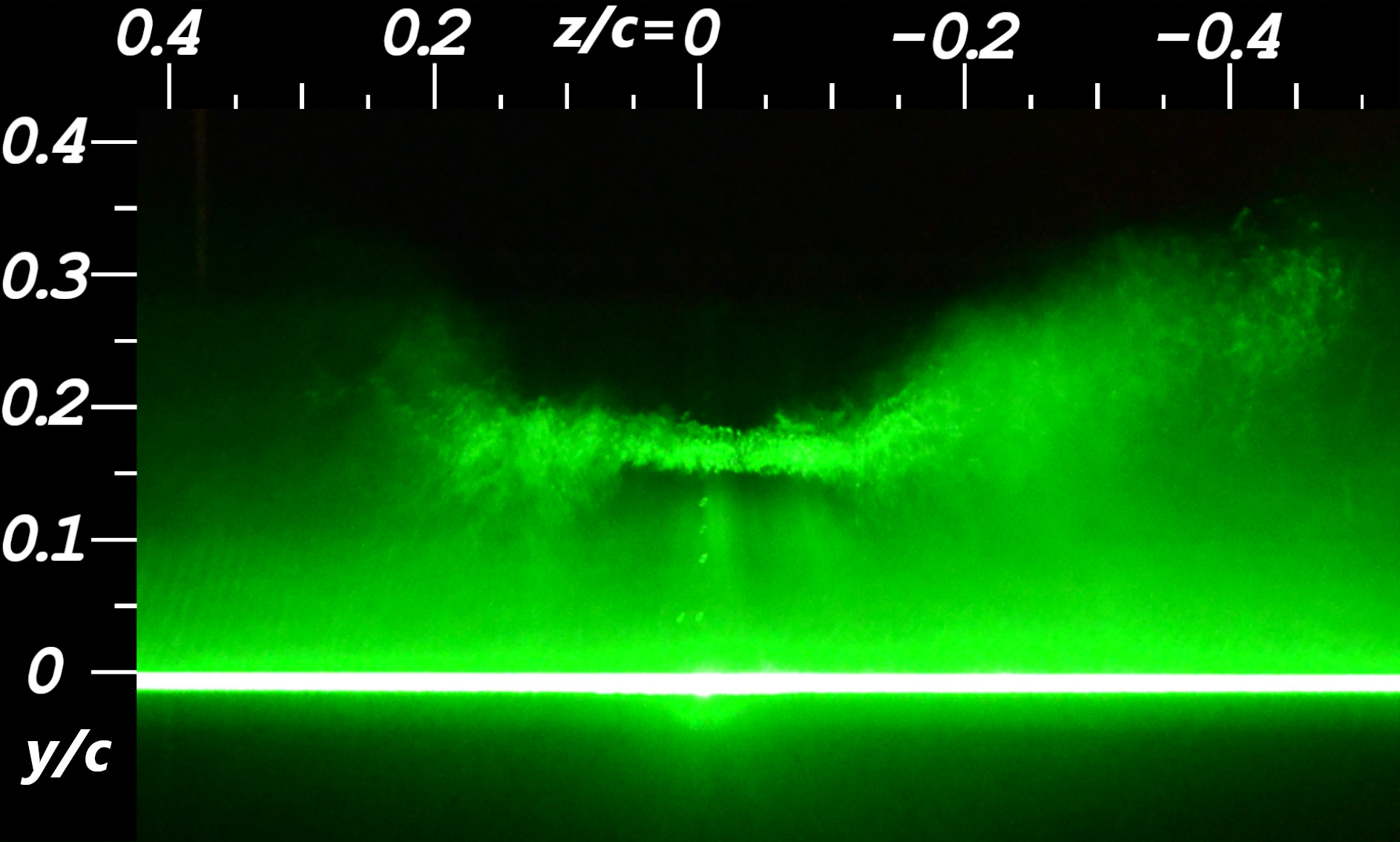}
        \caption{$C_B=5.0$; \SI{5}{\percent} DC}
        \label{fig:smoke_20V_5DC}
    \end{subfigure}
    \begin{subfigure}{0.302\linewidth}
        \includegraphics[width=\linewidth]{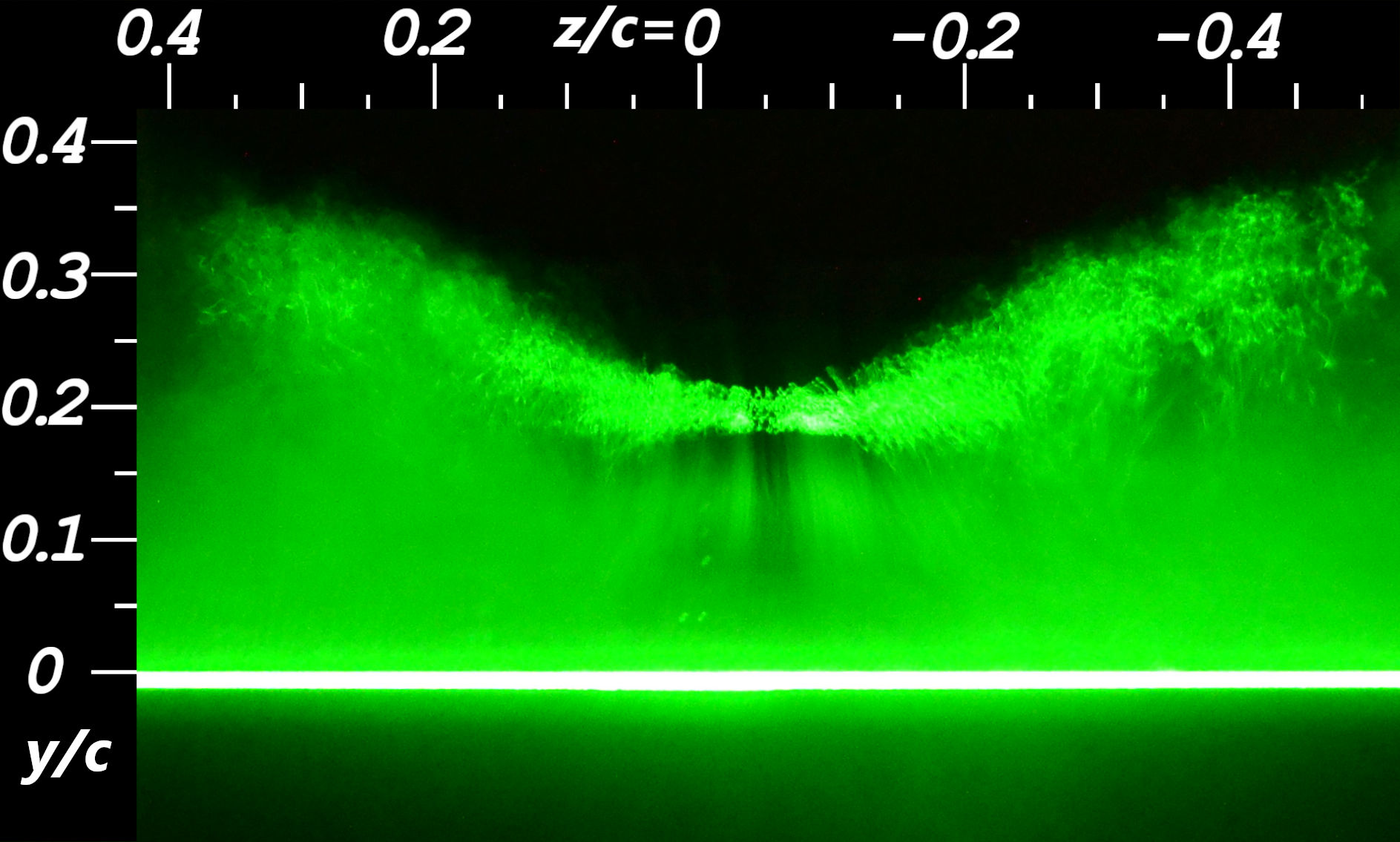}
        \caption{$C_B=1.9$; \SI{12.5}{\percent} DC}
        \label{fig:smoke_10V_12.5DC}
    \end{subfigure}
    \begin{subfigure}{0.302\linewidth}
        \includegraphics[width=\linewidth]{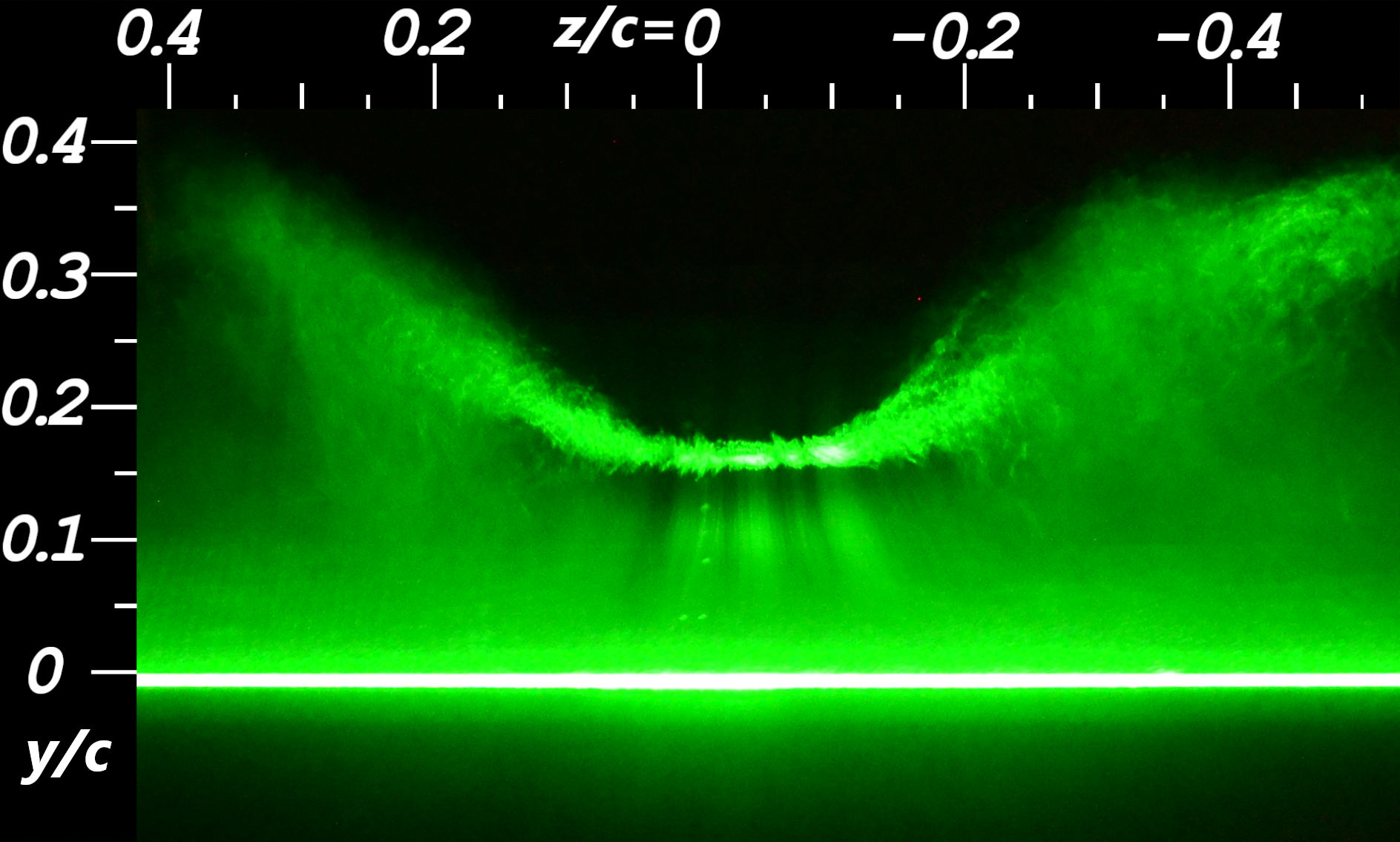}
        \caption{$C_B=4.4$; \SI{12.5}{\percent} DC}
        \label{fig:smoke_15V_12.5DC}
    \end{subfigure}
    \begin{subfigure}{0.302\linewidth}
        \includegraphics[width=\linewidth]{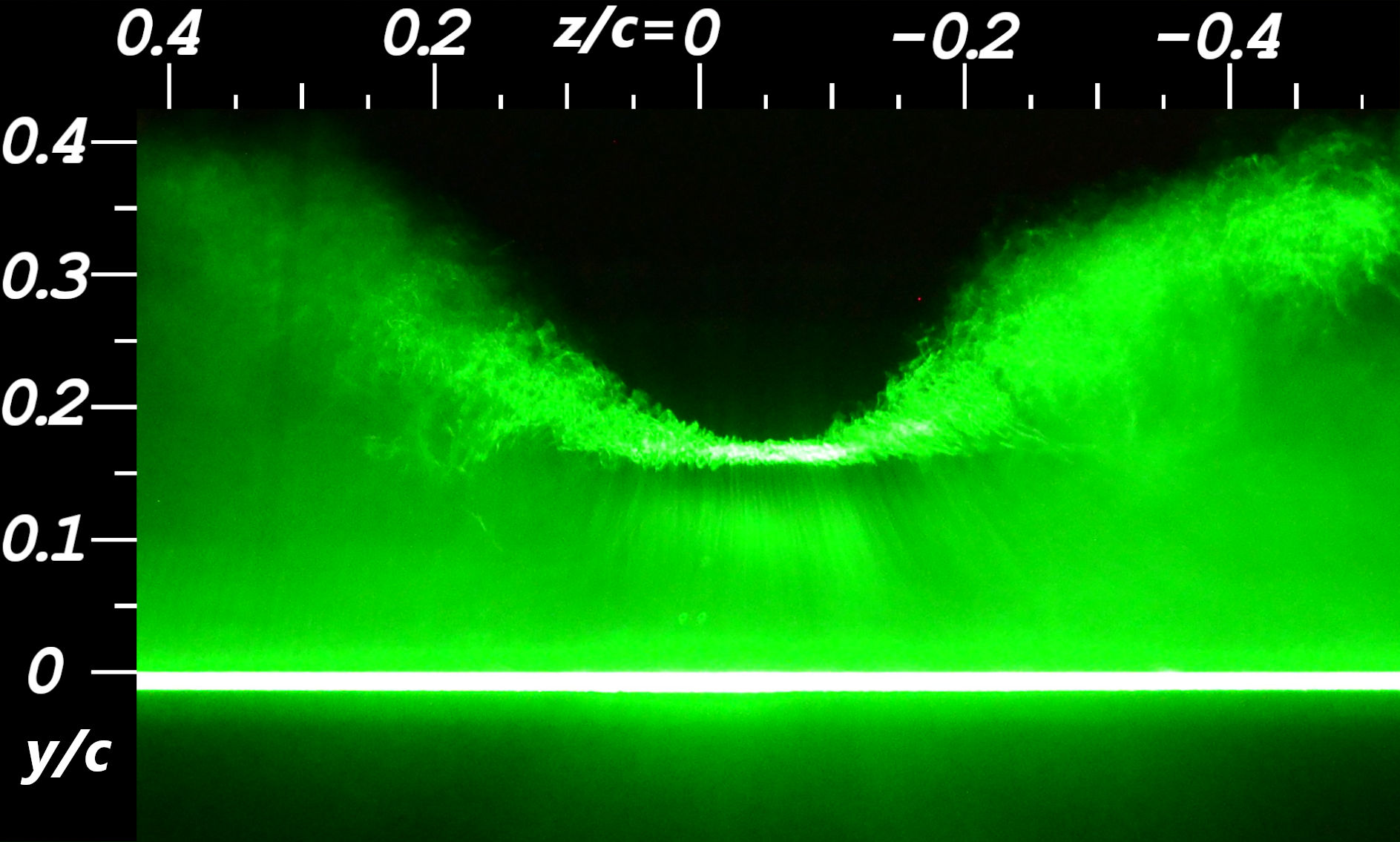}
        \caption{$C_B=5.0$; \SI{12.5}{\percent} DC}
        \label{fig:smoke_20V_12.5DC}
    \end{subfigure}
    \begin{subfigure}{0.302\linewidth}
        \includegraphics[width=\linewidth]{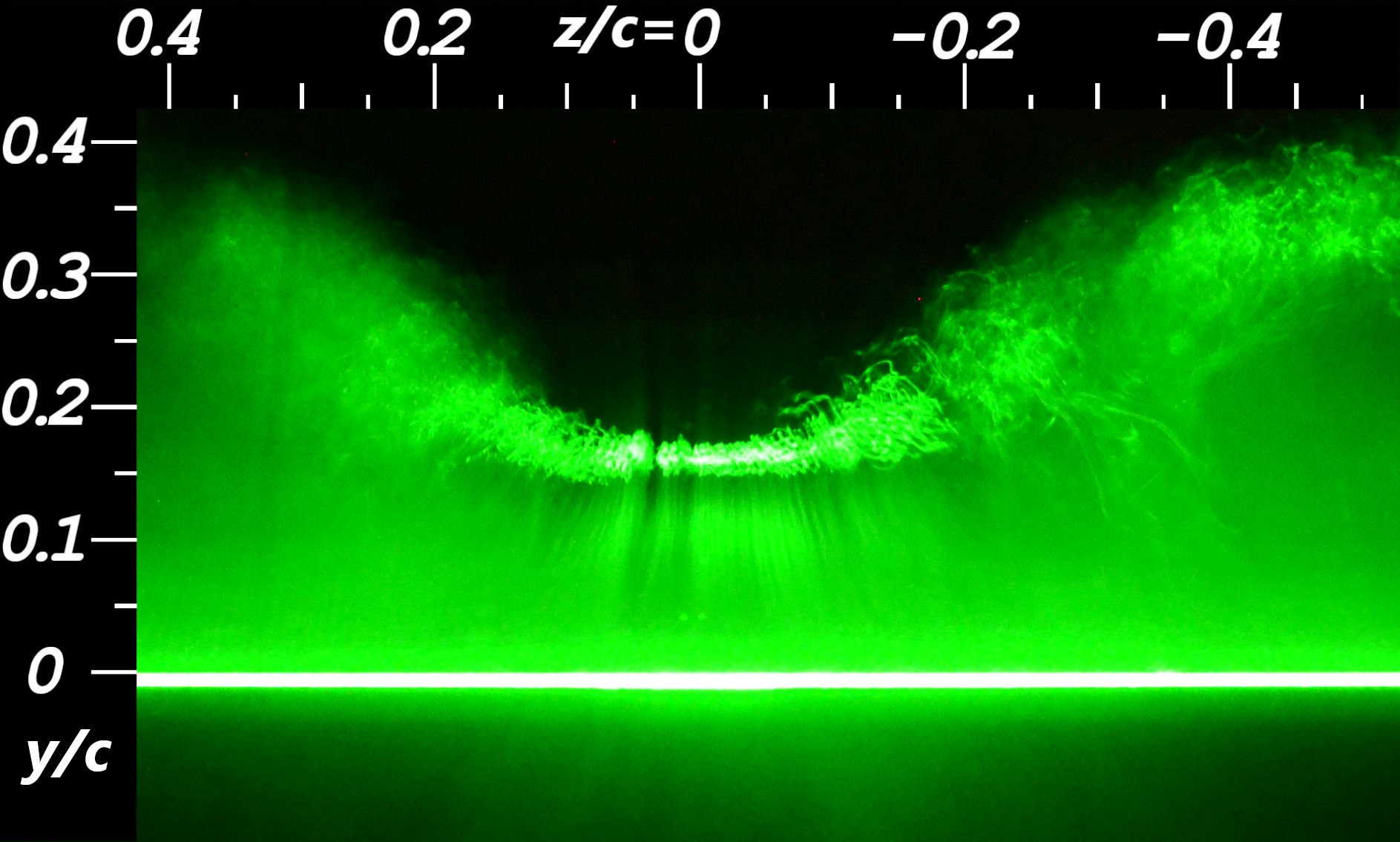}
        \caption{$C_B=1.9$; \SI{37.5}{\percent} DC}
        \label{fig:smoke_10V_37.5DC}
    \end{subfigure}
    \begin{subfigure}{0.302\linewidth}
        \includegraphics[width=\linewidth]{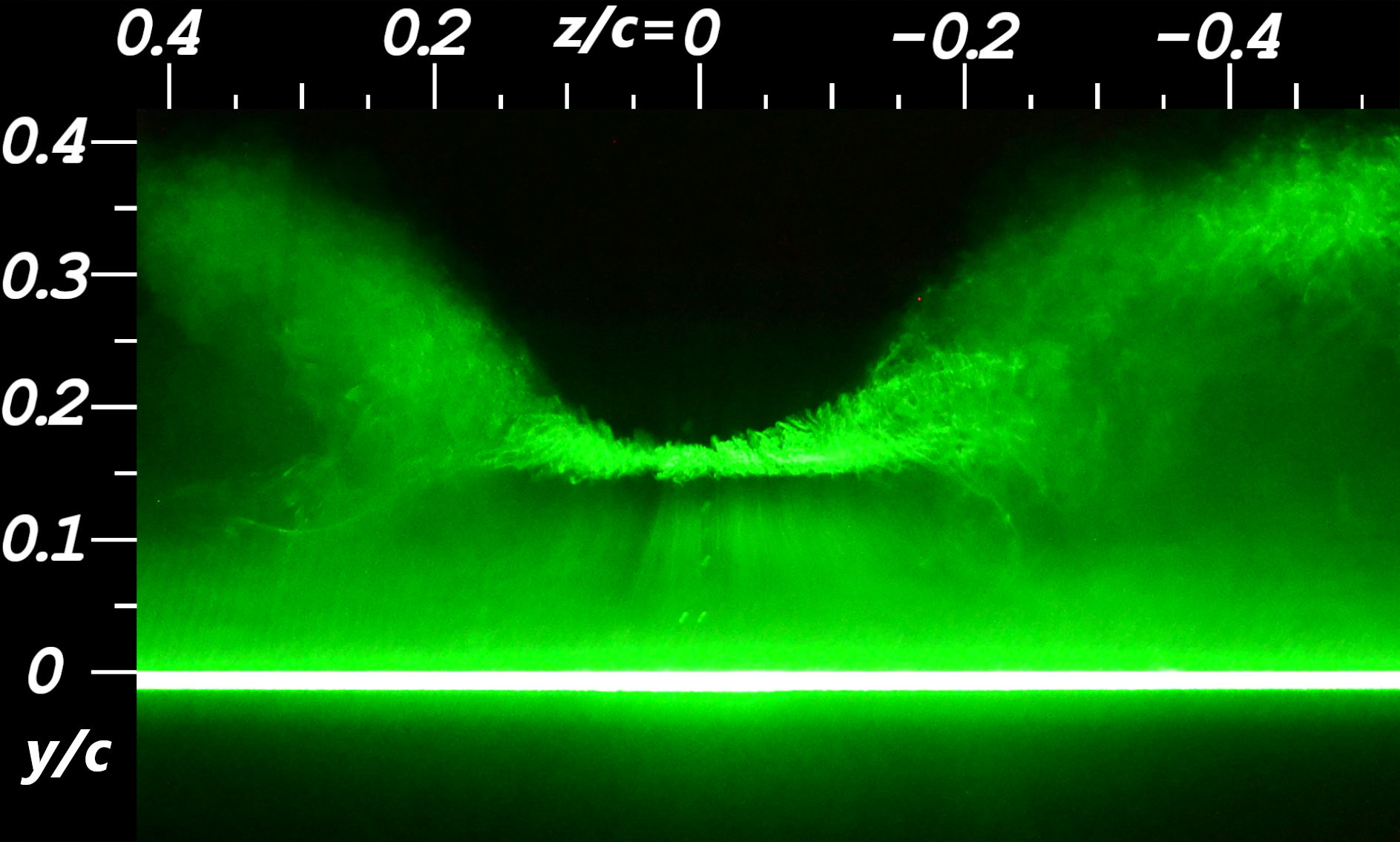}
        \caption{$C_B=4.4$; \SI{37.5}{\percent} DC}
        \label{fig:smoke_15V_37.5DC}
    \end{subfigure}
    \begin{subfigure}{0.302\linewidth}
        \includegraphics[width=\linewidth]{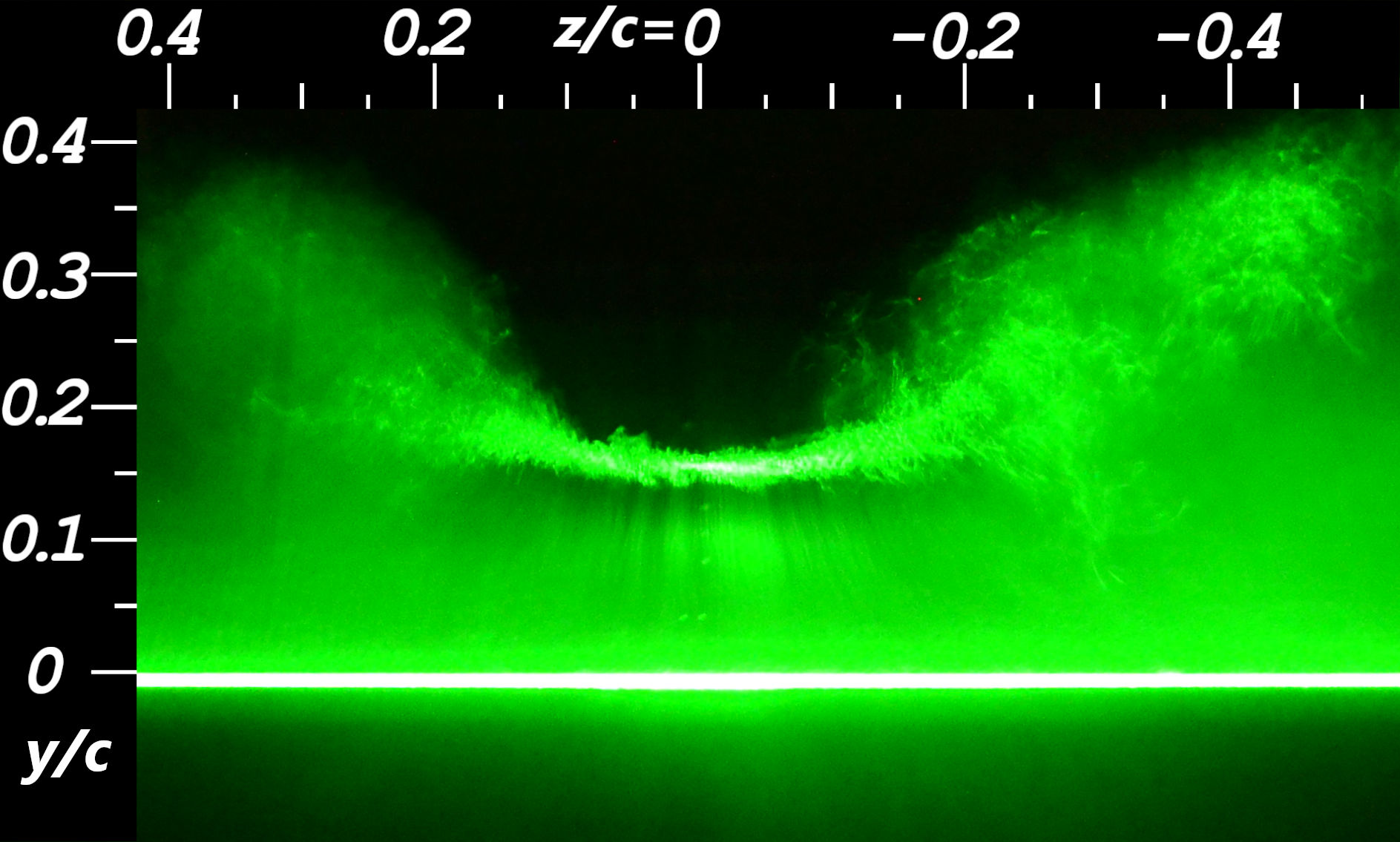}
        \caption{$C_B=5.0$; \SI{37.5}{\percent} DC}
        \label{fig:smoke_20V_37.5DC}
    \end{subfigure}
    \begin{subfigure}{0.302\linewidth}
        \includegraphics[width=\linewidth]{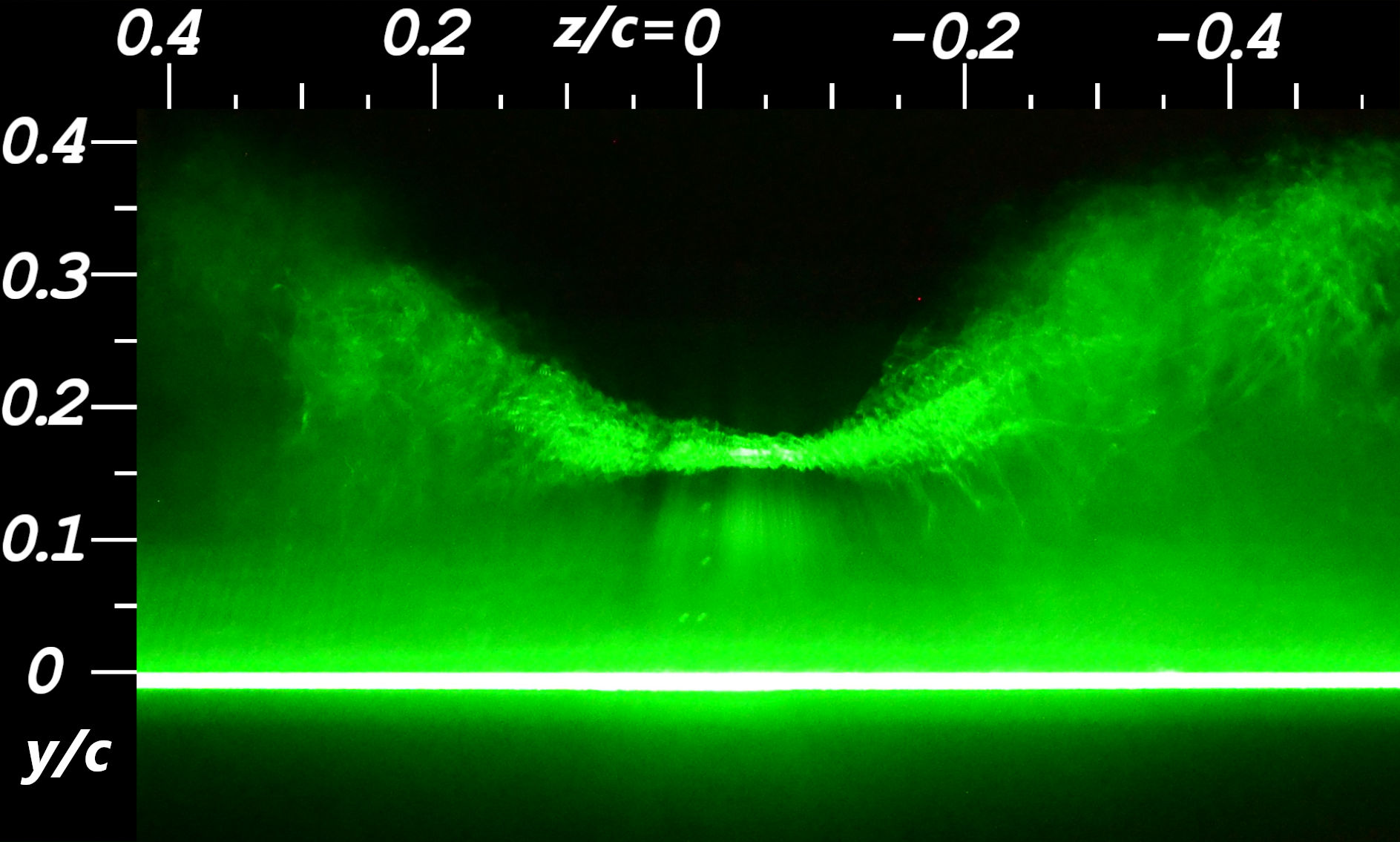}
        \caption{$C_B=1.9$; \SI{62.5}{\percent} DC}
        \label{fig:smoke_10V_62.5DC}
    \end{subfigure}
    \begin{subfigure}{0.302\linewidth}
        \includegraphics[width=\linewidth]{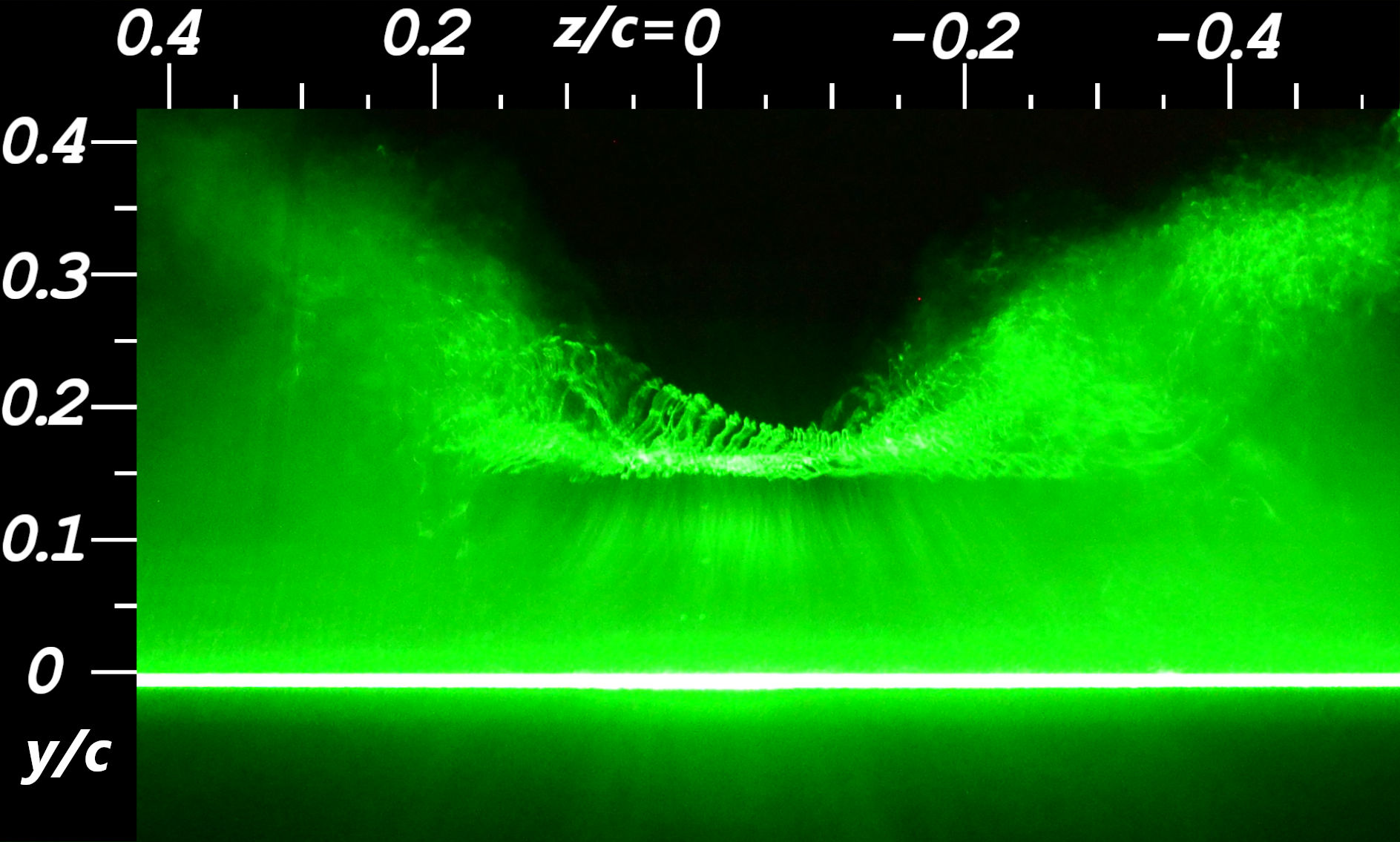}
        \caption{$C_B=4.4$; \SI{62.5}{\percent} DC}
        \label{fig:smoke_15V_62.5DC}
    \end{subfigure}
    \begin{subfigure}{0.302\linewidth}
        \includegraphics[width=\linewidth]{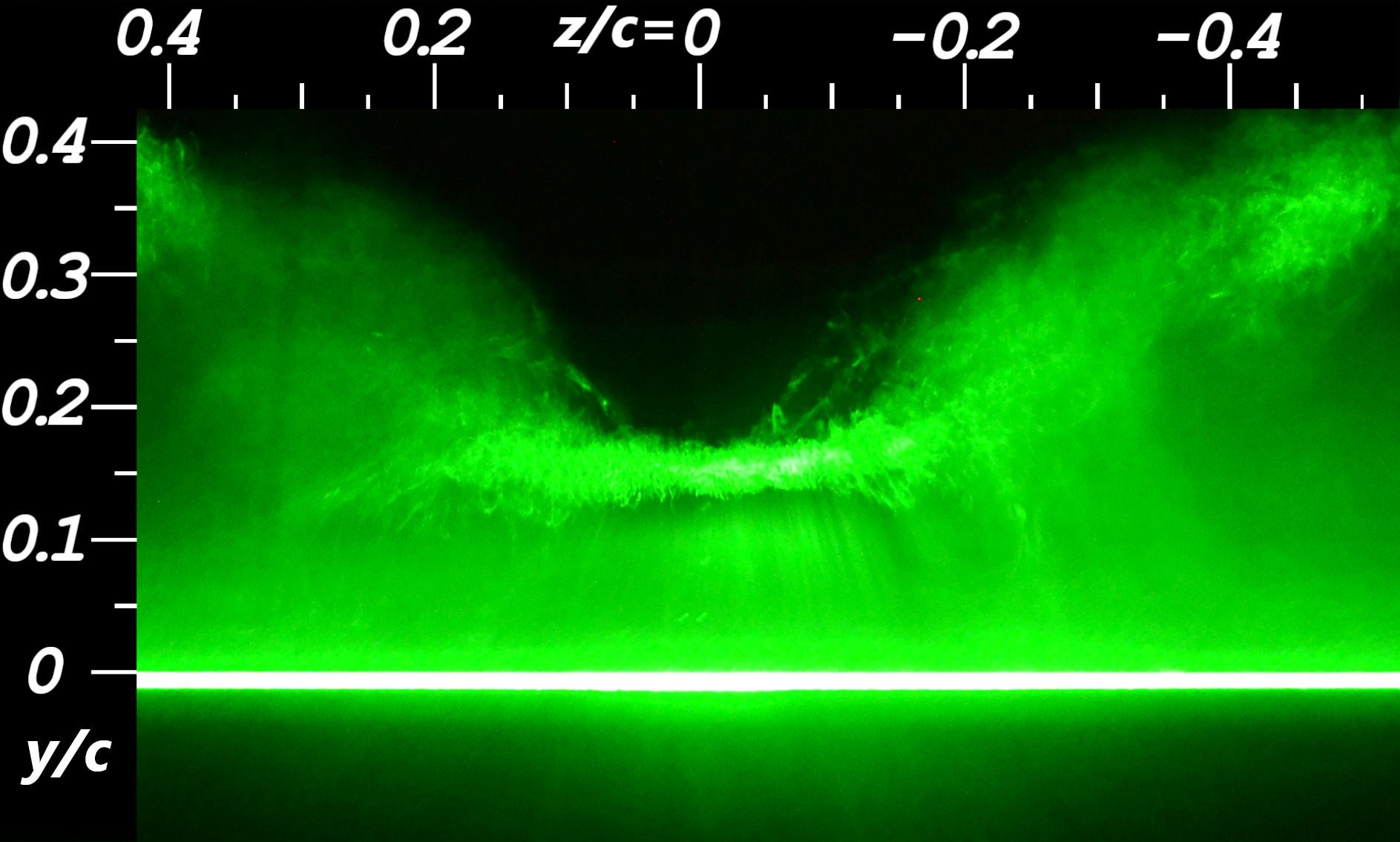}
        \caption{$C_B=5.0$; \SI{62.5}{\percent} DC}
        \label{fig:smoke_20V_62.5DC}
    \end{subfigure}
    \begin{subfigure}{0.302\linewidth}
        \includegraphics[width=\linewidth]{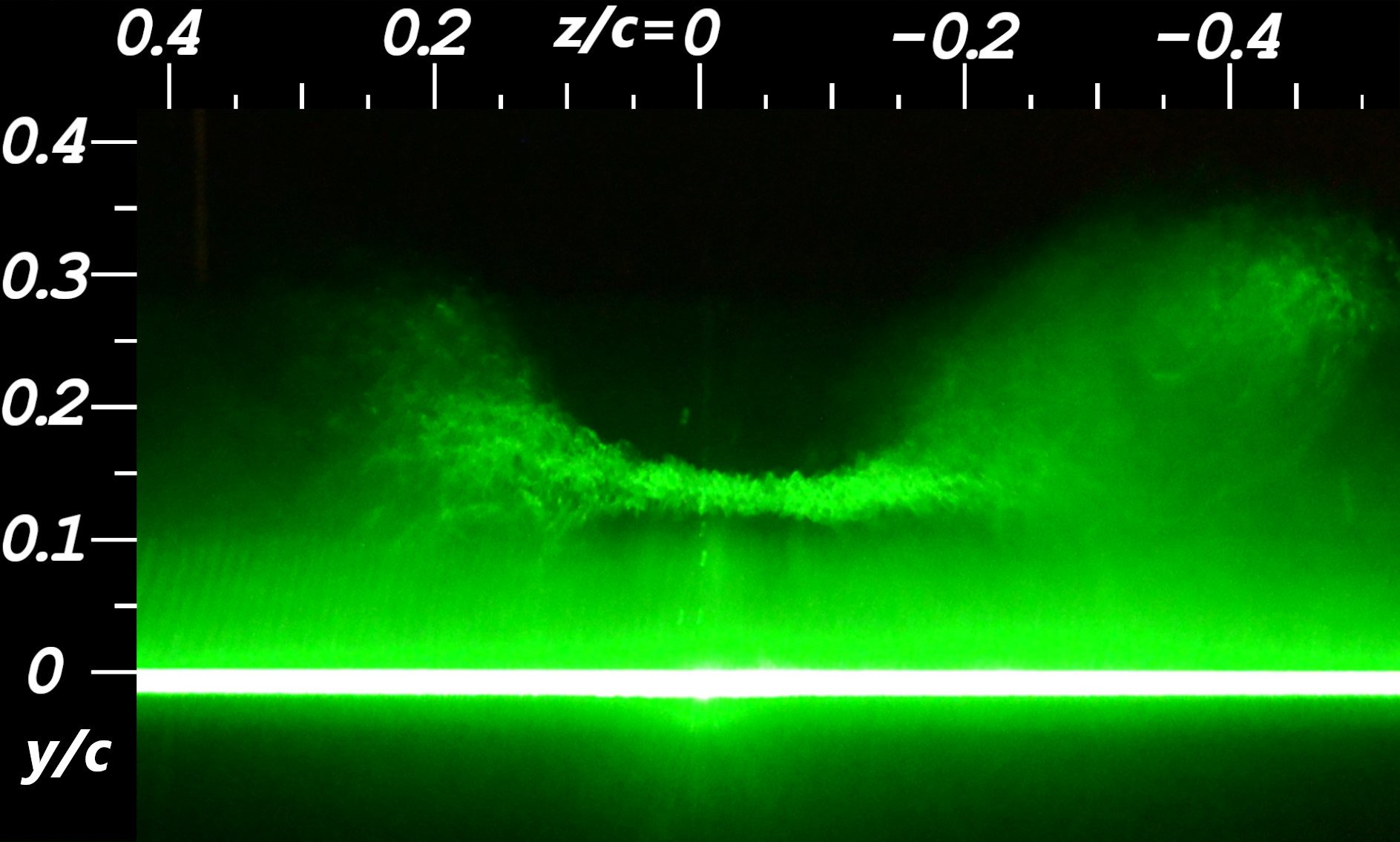}
        \caption{$C_B=1.9$; \SI{95}{\percent} DC}
        \label{fig:smoke_10V_95DC}
    \end{subfigure}
    \begin{subfigure}{0.302\linewidth}
        \includegraphics[width=\linewidth]{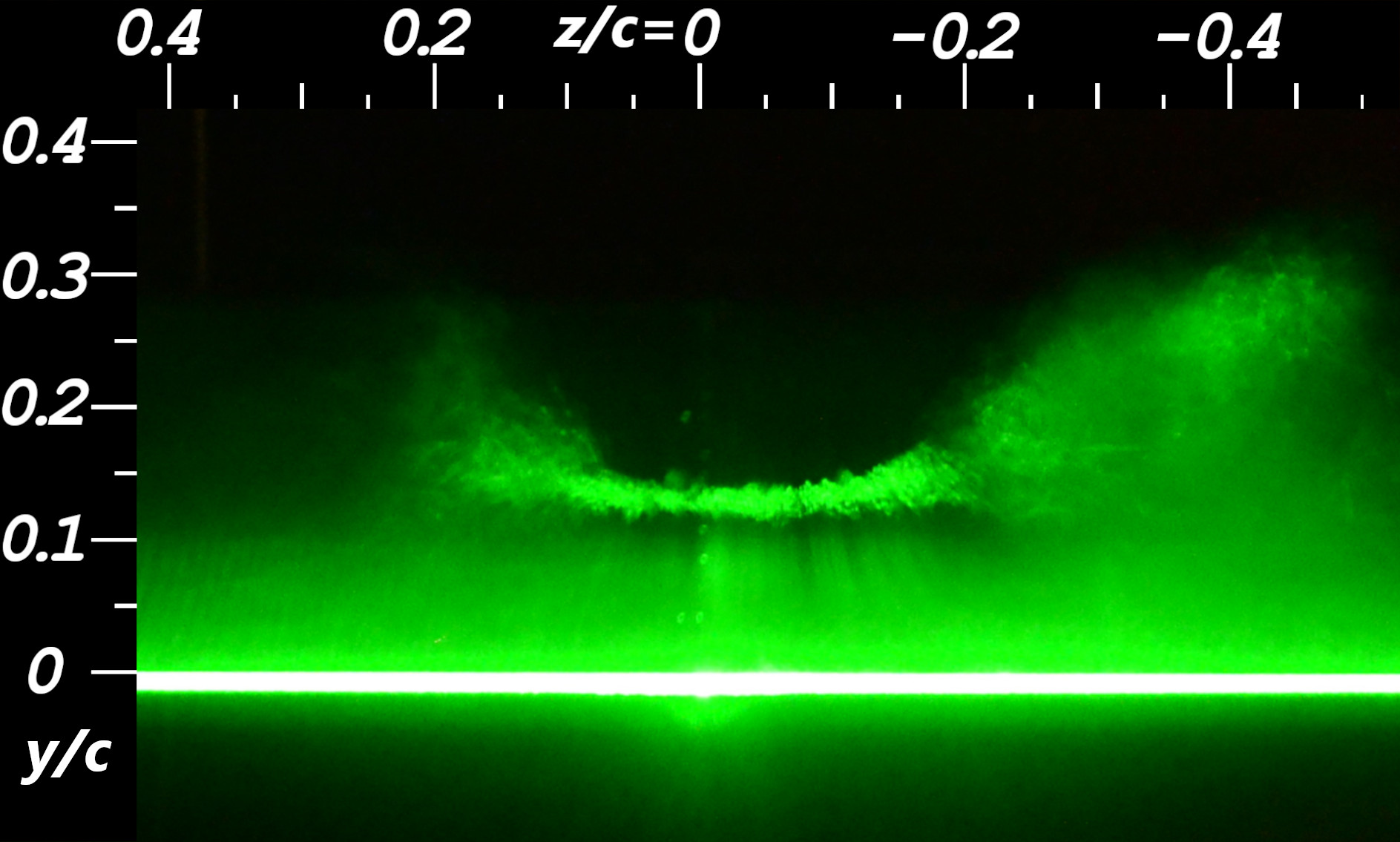}
        \caption{$C_B=4.4$; \SI{95}{\percent} DC}
        \label{fig:smoke_15V_95DC}
    \end{subfigure}
    \begin{subfigure}{0.302\linewidth}
        \includegraphics[width=\linewidth]{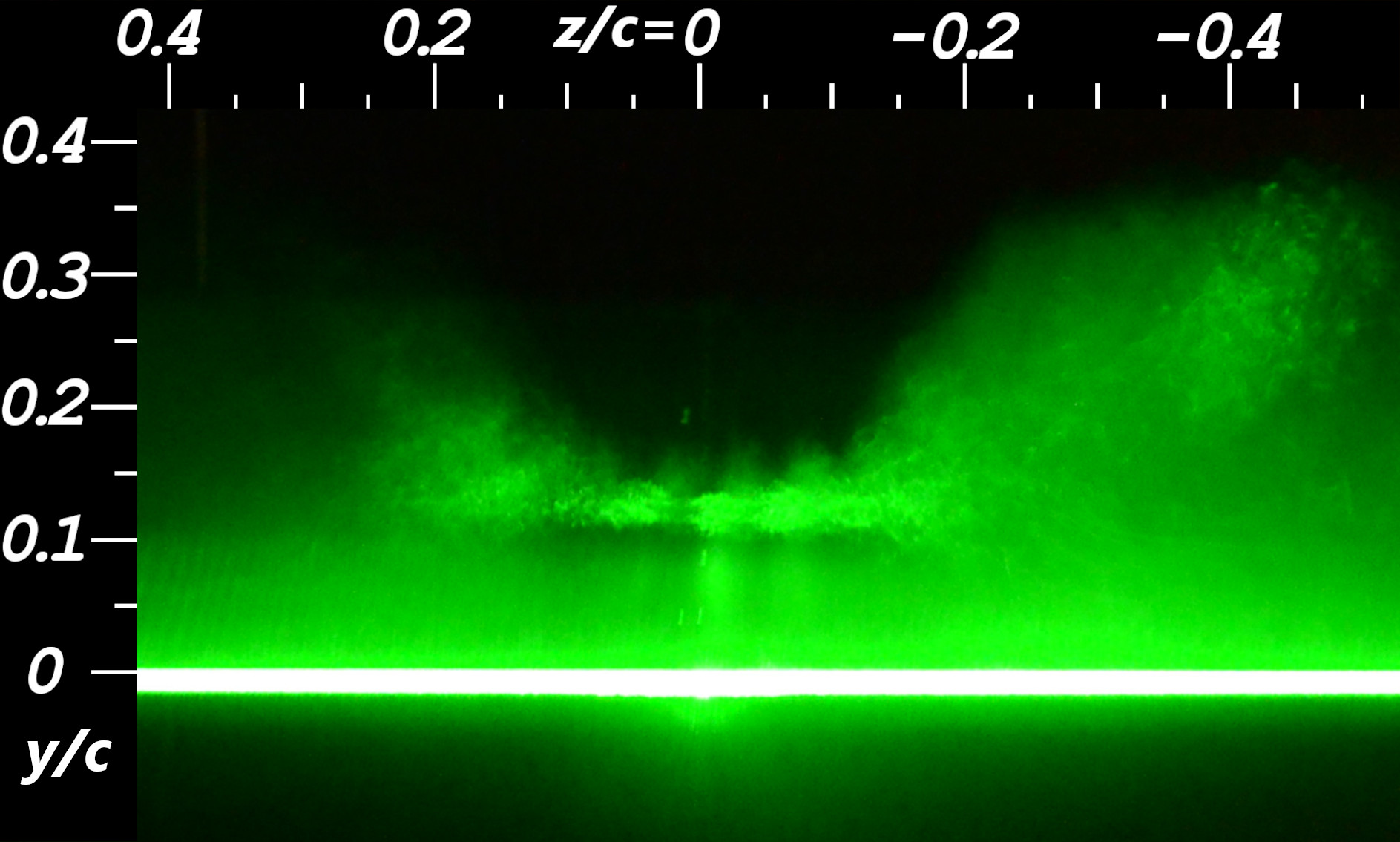}
        \caption{$C_B=5.0$; \SI{95}{\percent} DC}
        \label{fig:smoke_20V_95DC}
    \end{subfigure}
    \caption{Sectional smoke flow visualization at the trailing edge; spanwise-transverse plane}
    \label{fig:laser_smoke}
\end{figure}

To complement the pressure distributions, sectional flow visualizations at the trailing edge are presented in Figure~\ref{fig:laser_smoke} while providing additional insight into the spanwise control authority under a variety of actuation combinations. In the baseline stalled condition (Figure~\ref{fig:smoke_baseline}), a large area of diffuse smoke is observed above the airfoil, indicating separated and turbulent flow. At the lowest blowing ratio and duty cycle (Figure~\ref{fig:smoke_10V_5DC}), the SJA array does not produce sufficient momentum to reattach the separated shear layer, resulting in negligible flow control, which is consistent with the pressure plateau observed in Figure~\ref{fig:pressure_10V}. Asymmetric flow control is observed in the $C_B=4.4$ and \SI{5}{\percent} DC control case, with a consistent shift in the center of control toward the $-z$ direction across multiple smoke visualization runs. This repeatable behavior suggests that the asymmetry is not due to random variations in the flow but rather systematic differences in actuator performance. \citet{Xu2025} measured the spanwise velocity distribution across the same actuator array used in the present experiments, confirming jet-to-jet variations consistent with the asymmetry observed in the present control cases. These variations are attributed to minor variations in actuator fabrication, such as geometric tolerances or diaphragm stiffness, that can significantly affect the velocity response under low-voltage operation. The successful control cases (figures~\ref{fig:smoke_20V_5DC}~to~\ref{fig:smoke_20V_95DC}) exhibit a concentrated region of smoke streaks centered around the midspan corresponding to the reattached flow, with the streakline height representing the boundary layer thickness at the trailing edge. The effective spanwise control length for the reattached cases, marked by the dense smoke regions closest to the airfoil, is observed to be between $0.3c$ and $0.4c$, depending on the control case. Moving spanwise outward, the boundary layer thickens, forming the characteristic upward-concave shape observed in the visualized smoke. Beyond the midspan region ($|z/c| > 0.2$), the flow deteriorates into a turbulent, separated state with recirculation, as indicated by the diffuse smoke patterns. Previous studies have shown that the influence of the SJA array is largely confined to the midspan region, with the outer flow reverting to baseline conditions~\cite{Machado2024b,Machado2024a,Ho2024}. Figure~\ref{fig:smoke_10V_12.5DC} illustrates the effects of an overly power-conservative control strategy operating just above the reattachment threshold. Although the pressure distribution confirms flow reattachment, the boundary layer at midspan is noticeably thicker than in higher-power cases, as indicated by the presence of smoke streaks higher in the $y$-direction. Additionally, the spanwise control authority is diminished, with a noticeably shorter reattached region compared to higher momentum cases. Further momentum addition via increasing either the DC or blowing ratio results in a thinning of the boundary layer, corresponding to higher near-wall velocities, reduced wake thickness, and improved suction on the airfoil surface. The lengthened reattached region indicates improved spanwise control authority, aligned with prior oil flow visualizations showing that increased blowing ratios lead to a larger reattached flow region~\cite{Feero2017a}.

\begin{figure}
    \centering
    \includegraphics[width=\linewidth]{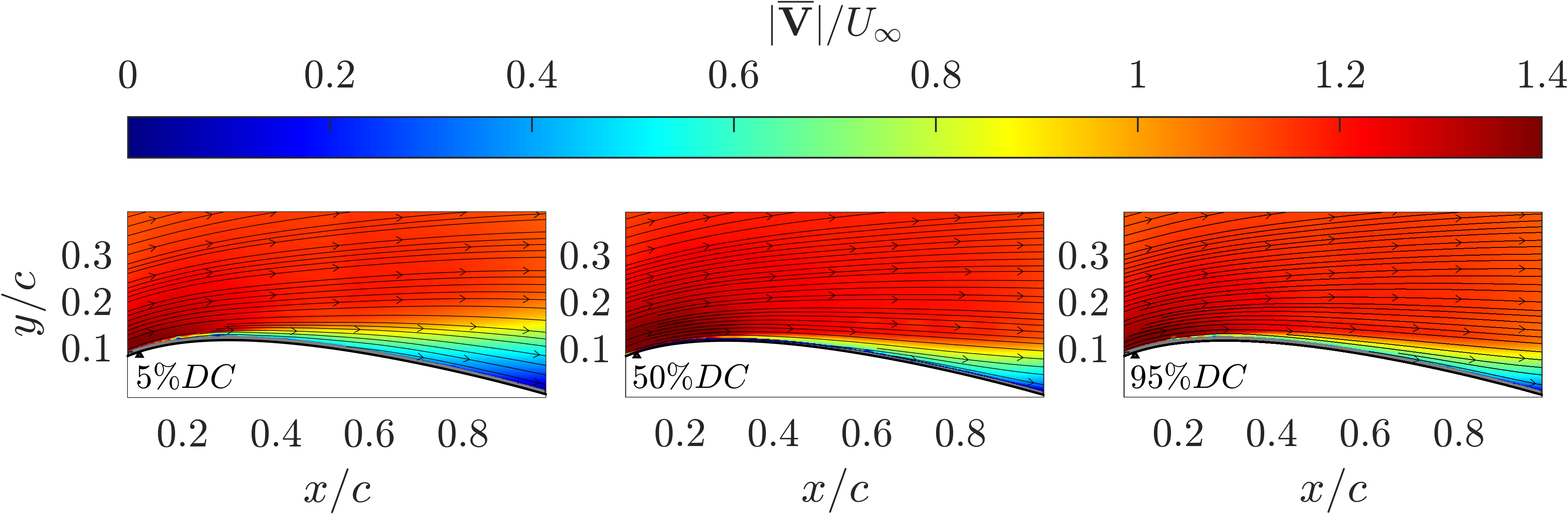}
    \caption{Streamlines and normalized mean velocity magnitude with control at $C_B=5.0$ and various DCs}
    \label{fig:velocity_magnitude}
\end{figure}

However, these improvements saturate at moderate power levels. The spanwise control length reaches a maximum at approximately \SI{40}{\percent} of the SJA array length, even with the highest power strategies tested. This limit arises from the highly turbulent, uncontrolled region beyond the extent of the array, which disrupts the controlled flow~\cite{Machado2024b}. Similarly, the smoke visualizations suggest that boundary layer thinning shows diminishing returns beyond moderate DC and blowing ratios. Contours of the time-averaged velocity magnitude with streamlines overlaid are presented in Figure~\ref{fig:velocity_magnitude}, confirming this observation. The position of the SJA array is marked as a black triangle at $x/c=0.1$. As the DC increases from \SI{5}{\percent} to \SI{50}{\percent}, the boundary layer becomes thinner, indicating reduced drag and aerodynamic improvements. However, increasing the DC to \SI{95}{\percent} yields no further boundary layer reduction, with mean flow characteristics remaining largely unchanged from the \SI{50}{\percent} case. These observations indicate that while increasing momentum input improves control effectiveness beyond the reattachment threshold, practical saturation limits exist for both spanwise control authority and boundary layer thinning, suggesting that more efficient control methods with lower power requirements are achievable.

\begin{figure}
    \begin{subfigure}{0.483\linewidth}
        \includegraphics[width=\linewidth]{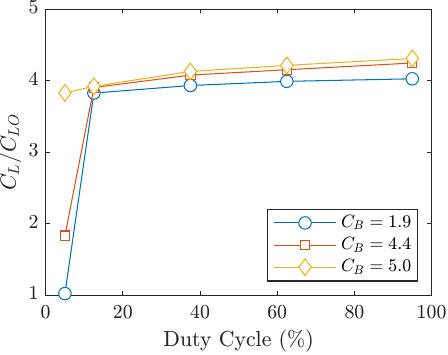}
        \caption{Normalized lift coefficient variation with DC}
        \label{fig:lift_vs_DC}
    \end{subfigure}
    \hfill
    \begin{subfigure}{0.497\linewidth}
        \includegraphics[width=\linewidth]{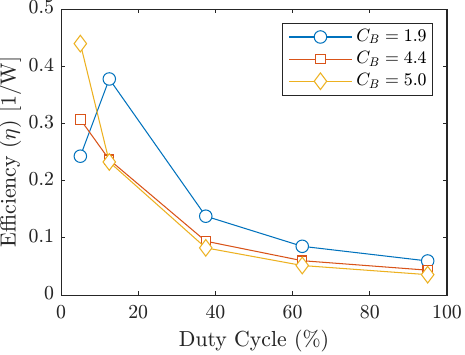}
        \caption{Control efficiency variation with DC}
        \label{fig:lift_efficiency}
    \end{subfigure}
    \begin{subfigure}{0.49\linewidth}
        \includegraphics[width=\linewidth]{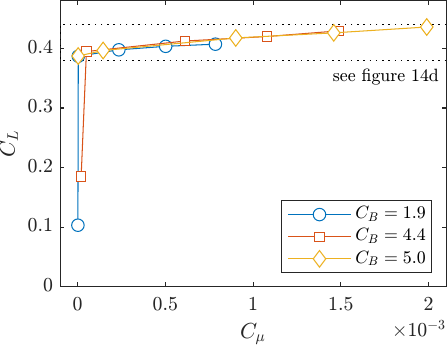}
        \caption{Lift variation with momentum coefficient \newline}
        \label{fig:lift_momentum}
    \end{subfigure}
    \hfill
    \begin{subfigure}{0.49\linewidth}
        \includegraphics[width=\linewidth]{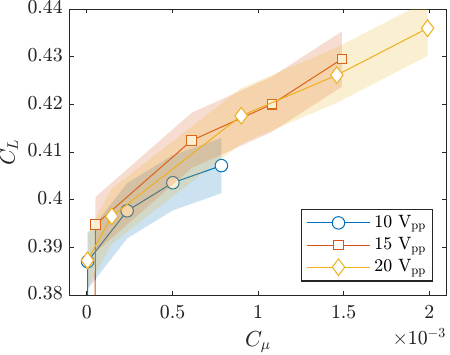}
        \caption{Zoomed in lift variation with momentum coefficient. The highlighted regions indicate the \SI{95}{\percent} CI for each plot}
        \label{fig:lift_momentum_zoomed}
    \end{subfigure}
    \caption{Midspan lift characteristics across varying DCs, blowing ratios, and momentum coefficients}
    \label{fig:lift_plots}
\end{figure}

These saturation effects are further evident when examining lift characteristics across the parameter space. Figure~\ref{fig:lift_vs_DC} presents midspan lift coefficients normalized by the baseline lift coefficient for various control strategies. The lift coefficients presented reflect midspan measurements only and do not capture the three-dimensional variations shown in the smoke visualizations. The measured baseline lift coefficient, $C_{LO}=0.101\pm0.005$, is consistent with prior studies with the same airfoil model~\cite{Feero2017b,Xu2023}. The curves for the lower two blowing ratios exhibit a sharp increase in lift up to a threshold level, corresponding to fully attached flow. Beyond this threshold, further increases in DC past \SI{12.5}{\percent} result in only marginal improvements in lift. Similarly, at a constant DC, increasing the blowing ratio leads to higher lift coefficients. A large lift increase of \SI{280}{\percent} over the baseline case is achieved at a low power consumption with $C_B=1.9$, \SI{12.5}{\percent} DC. However, only an additional \SI{40}{\percent} lift increase is attained at the highest power control strategy, which requires 12 times the power input.

To quantify the tradeoff between power input and aerodynamic improvements, lift coefficients are expressed relative to the power consumption for each control strategy, resulting in a measure of efficiency,
\begin{equation}
    \eta=\frac{C_L}{P_\mathrm{SJA}}.
\end{equation}
This efficiency metric is plotted against the DC in Figure~\ref{fig:lift_efficiency}. The plot reveals that once the flow is fully reattached, the lift-to-power efficiency of the control decreases as either the blowing ratio or DC is increased. Furthermore, the highest efficiency is achieved with the highest blowing ratio combined with the lowest DC. This suggests that strong, yet brief perturbations to the oncoming flow are sufficient for time-averaged reattachment while maintaining power efficiency. Additionally, at sufficiently high DCs, reducing the blowing ratio results in power savings with minimal impact on lift performance.

To examine the interactions between the DC and blowing ratio on the momentum coefficient and their combined effect on lift, the lift coefficient response to varying momentum coefficients is presented in Figure~\ref{fig:lift_momentum}. The zoomed-in view in Figure~\ref{fig:lift_momentum_zoomed} reveals that these curves overlap within experimental uncertainty. This demonstrates that a given momentum coefficient produces consistent lift for reattached flows, regardless of the specific DC-blowing ratio combination used to achieve it. This confirms that the momentum coefficient is the governing parameter for time-averaged aerodynamic performance, consistent with established flow control principles~\cite{Seifert1999,Greenblatt2000,Goodfellow2013}.

\subsection{Effect of Duty Cycle on Flow Stability}
\label{stability}
\begin{figure}
    \centering
    \includegraphics[width=0.6\linewidth]{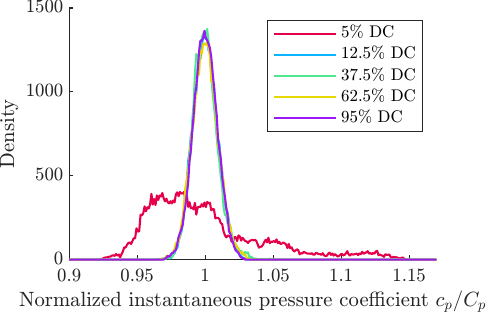}
    \caption{Distributions of instantaneous pressure coefficients at the suction peak ($x/c=0.07$) for control at $C_B=4.4$, normalized by the mean coefficient of each DC}
    \label{fig:CP_histogram}
\end{figure}

\begin{figure}
    \centering
    \includegraphics[width=0.7\linewidth]{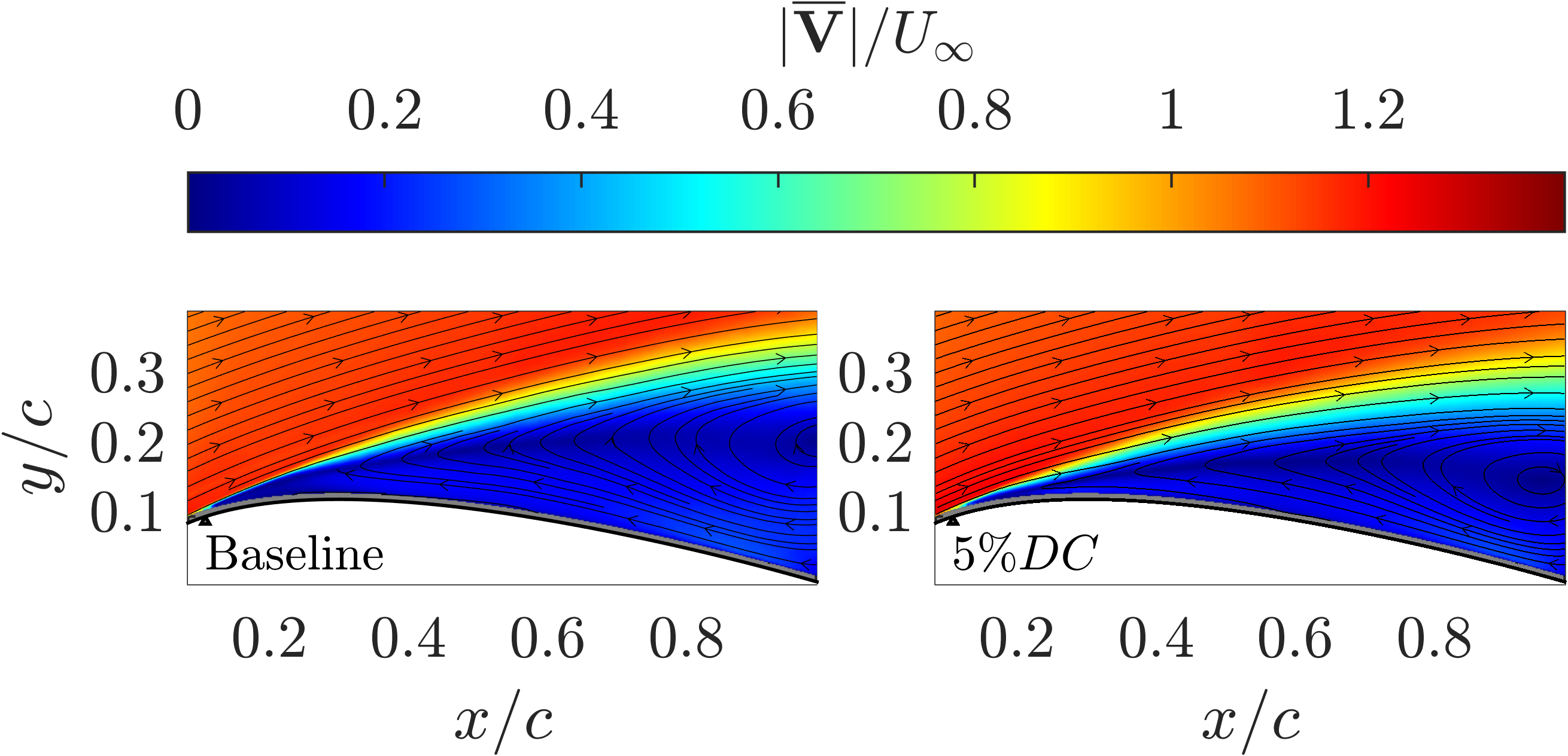}
    \caption{Streamlines and normalized mean velocity magnitude for the baseline case, and with control at $C_B=4.4$ and \SI{5}{\percent} DC}
    \label{fig:velocity_mag_15V_5DC}
\end{figure}

Distributions of normalized instantaneous pressure coefficients measured at the suction peak ($x/c=0.07$) are presented in Figure~\ref{fig:CP_histogram} for control at $C_B=4.4$ across various DCs. For DCs $\geq$ \SI{12.5}{\percent}, the pressure measurements are normally distributed and narrow, indicating a consistent suction peak and stable lift. In contrast, the \SI{5}{\percent} DC case exhibits a wider, right-skewed distribution (toward the higher suction side), indicating unsteady control and intermittent shear layer flapping. This low-power control case induces a transitional state between a fully attached flow and a massively separated flow, consistent with previous observations for weak SJA actuation~\cite{Tang2014, Salunkhe2016}. The moderate energy input does not fully eliminate flow separation but does improve the flow field by reducing the severity of the separation. This is further evidenced by the plots of mean velocity magnitude shown in Figure~\ref{fig:velocity_mag_15V_5DC}. In both the baseline and controlled case, the flow separates at $x/c=0.12$, consistent with prior findings~\cite{Xu2023}, forming a large mean recirculation region. In the baseline case, the core of the recirculation region is centered about the trailing edge, whereas in the mildly controlled case, the core shifts slightly upstream, and the separated shear layer becomes smaller.

\begin{figure}
    \centering
    \includegraphics[width=\linewidth]{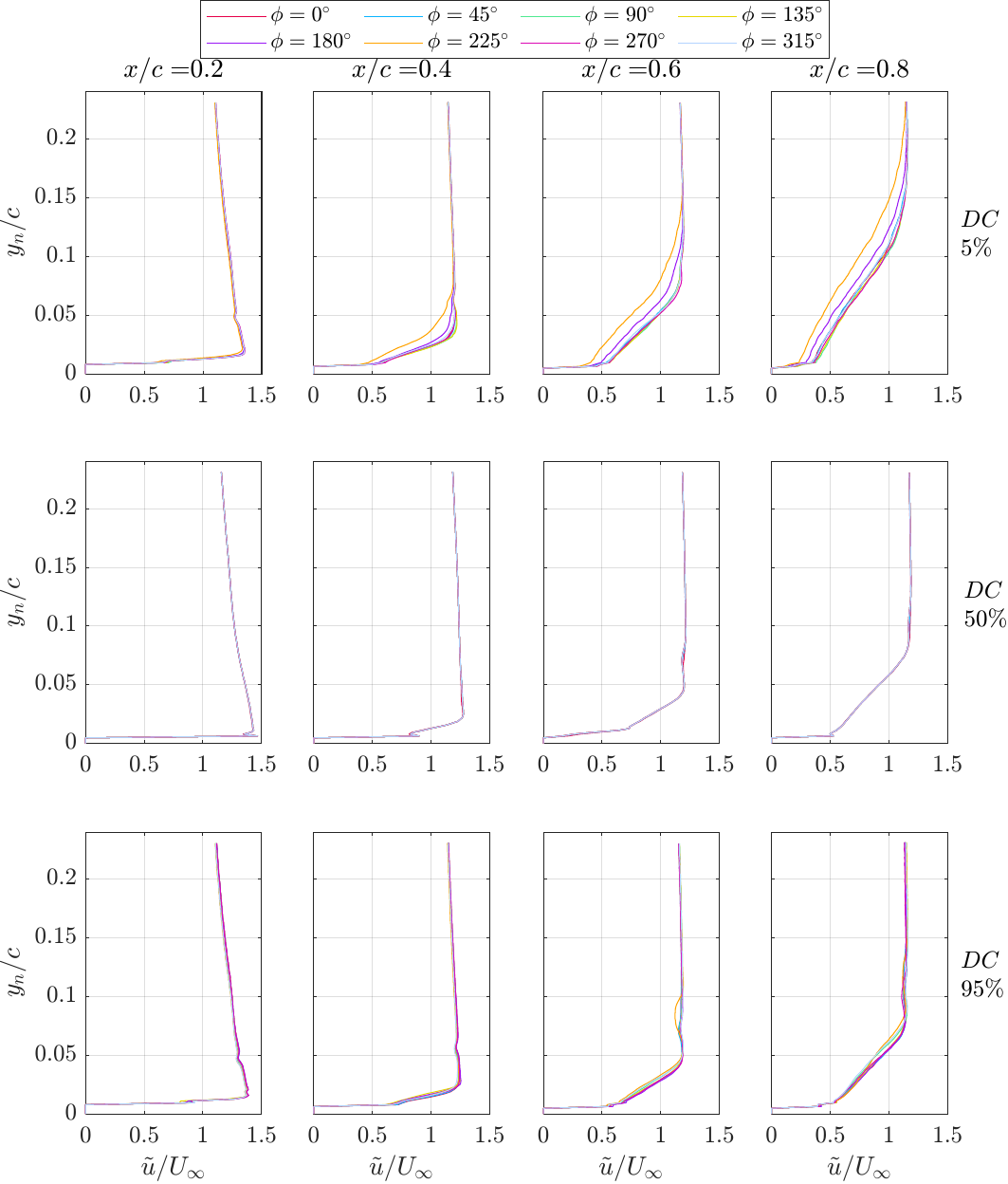}
    \caption{Wall-normal profiles of streamwise phase velocity along the chord with control at $C_B=5.0$ and various DCs}
    \label{fig:phase_velocity_profiles}
\end{figure}

Wall-normal, phase-averaged velocity plots at various chordwise stations are presented in Figure~\ref{fig:phase_velocity_profiles}. These plots show the phase-averaged streamwise velocity normalized by the freestream velocity as a function of wall-normal distance. Each curve represents a different phase within the actuation cycle. At a \SI{5}{\percent} DC, the velocity profiles exhibit substantial variation across different phase angles, indicating an unsteady and phase-dependent flow. In contrast, at higher duty cycles (\SI{50}{\percent} and above), the profiles become quasi-steady with minimal phase-to-phase variation, suggesting greater flow stability. The increased flow stability at higher duty cycles is also evident in the smoke visualizations (Figure~\ref{fig:laser_smoke}), where the streaks in the reattached region exhibit a tighter spread during the long-exposure image. At low DCs, the near-wall velocity remains low, consistent with the measured reduction in lift and the presence of a thicker boundary layer. Increasing the DC to \SI{50}{\percent} raises the near-wall velocity and thins the boundary layer, indicating enhanced performance. Further increasing the duty cycle to \SI{95}{\percent} produces no noticeable improvement in the velocity profiles and, in fact, introduces slight unsteadiness, as evidenced by the broader spread in the phase-averaged velocity profiles. The small velocity deficits observed above the boundary layer at $y_n/c=0.07$ correspond to the cores of streamwise vortex rings induced by the SJA~\cite{Xu2023,Machado2024b}.

\subsection{Effects of Duty Cycle on Vortical Structures}
\label{sec:duty_cycle}
\begin{figure}
    \centering
    \includegraphics[width=\linewidth]{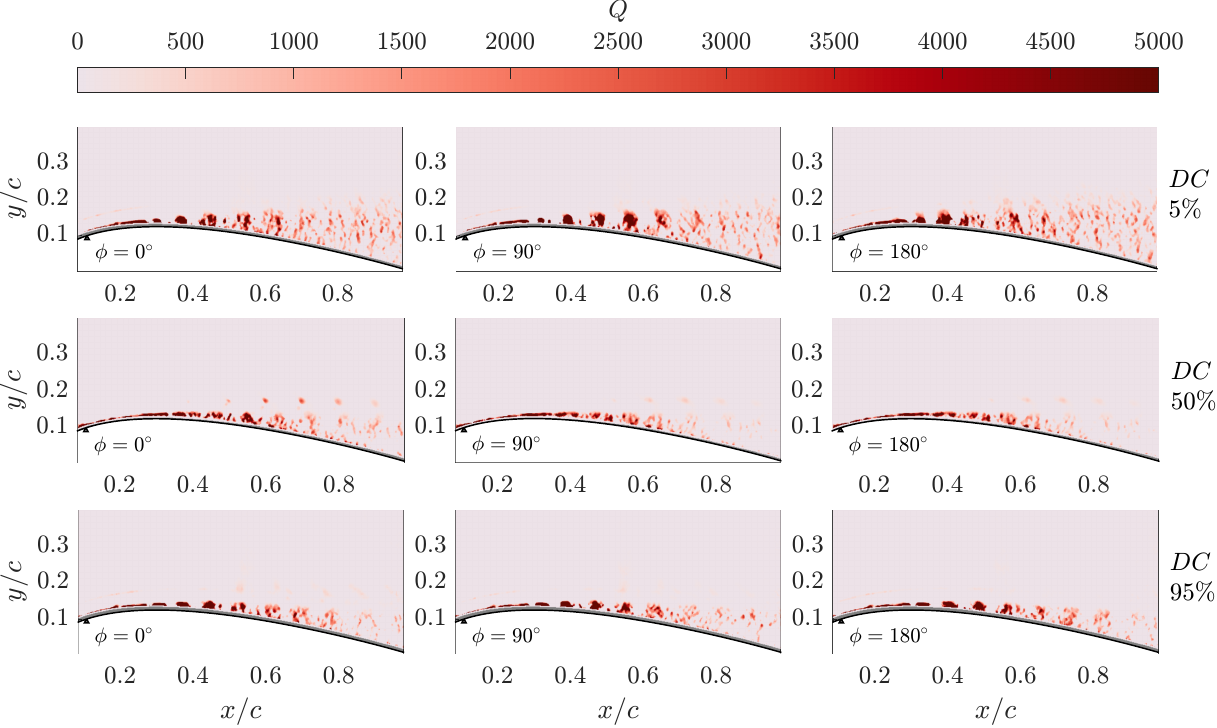}
    \caption{Evolution of vortex structures across the actuation period with control at $C_B=5.0$ and various DCs}
    \label{fig:Q_contour}
\end{figure}

\begin{figure}
    \centering
    \includegraphics[width=\linewidth]{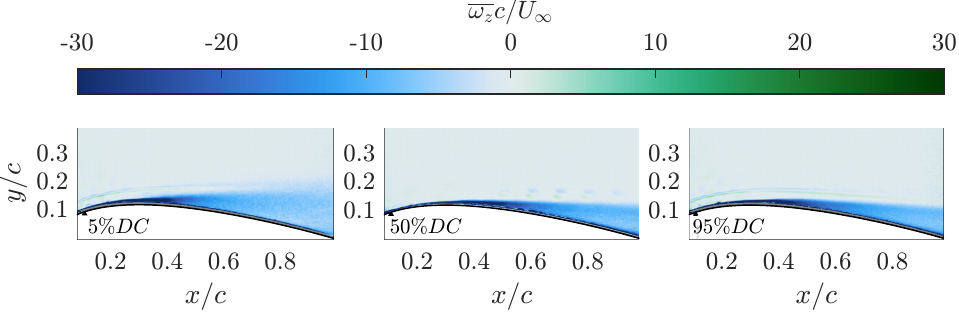}
    \caption{Non-dimensional vorticity with control at $C_B=5.0$ and various DCs}
    \label{fig:vorticity}
\end{figure}

The $Q$-criterion, defined as the second invariant of the velocity gradient tensor, is used to identify vortical structures in the controlled flow fields. For the 2D velocity field, $Q$ is defined as:
\begin{equation}
    Q=\frac{1}{2}(|\omega_z|^2-S^2),
\end{equation}
where $\omega_z$ is the out-of-plane vorticity, and $S$ is the strain-rate tensor~\cite{Hunt1998}. Contours of the $Q$-criterion are plotted in Figure~\ref{fig:Q_contour} across three phase angles, illustrating the evolution of vortical structures during the actuation period. The vortices exhibit a clockwise rotational sense, as indicated by the vorticity field (Figure~\ref{fig:vorticity}), which aligns with the typical behavior of SJA-induced vortices. With a \SI{5}{\percent} DC, larger vortex structures are observed, which tend to lift off the airfoil surface and dissipate more quickly, reducing their effectiveness. In contrast, higher DCs produce smaller, stronger, and more persistent vortices that remain closer to the wall, effectively energizing the shear layer. The observed vortex configuration provides the physical mechanism for the noted flow acceleration and lift enhancement, consistent with prior findings that link post-reattachment control improvements to stronger, more persistent vortices generated by enhanced jet injection~\cite{Liu2022,Yang2022}.

High-frequency actuation typically produces quasi-steady control by generating new vortices before the preceding vortices have convected over the airfoil~\cite{Amitay2002a, Glezer2005,feero2015a, Kim2022,Xu2023,Machado2024b,Xu2025effect}. This occurs when the actuation period, $T = 1/f_m$, is much shorter than the convective timescale, $t_c = c/U_{\infty}$; in the present study, $T \approx 0.1t_c$. However, low-DC control disrupts this quasi-steady behavior, introducing significant phase-dependence. At \SI{5}{\percent} DC, the chordwise dissipation point of vortices varies throughout the actuation cycle. At $\phi=\ang{90}$, a train of vortices persists until $x/c = 0.6$ before dissipating, whereas at $\phi=\ang{180}$, dissipation occurs farther upstream (around $x/c = 0.4$). The dissipation of vortices farther upstream reduces momentum transport and results in slower near-wall velocities. The phase velocity profiles in Figure~\ref{fig:phase_velocity_profiles} exhibit velocity deficits at $\ang{180}$ and $\ang{225}$ for $x/c \geq 0.4$, corresponding precisely to the phases and chordwise locations where vortex dissipation occurs. This correspondence demonstrates that phase-dependent vortex dissipation governs flow unsteadiness at low DCs. At higher DCs, more consistent SJA actuation produces coherent vortices across phases, leading to the tighter spread of phase velocity profiles observed in the \SI{50}{\percent} and \SI{95}{\percent} DC cases.

Time-averaged vorticity plots (Figure~\ref{fig:vorticity}) complement the phase-averaged $Q$-criterion contours, with negative vorticity above the airfoil indicating clockwise rotation of the spanwise vortices. At a low DC of \SI{5}{\percent}, vorticity spreads farther from the surface, reflecting rapid dissipation and reduced momentum transport. In contrast, at higher DCs of \SI{50}{\percent} and \SI{95}{\percent} DC, vorticity remains concentrated near the airfoil surface, indicating stronger vortices that more effectively energize the near-wall flow. This concentration of vorticity corresponds to the previously discussed thinner boundary layer, enhanced circulation, and increased lift. Additionally, the high-vorticity region ($\overline{\omega_z}c/U_\infty>30$) extends farther downstream at higher DCs, consistent with more persistent vortices, whereas at \SI{5}{\percent} DC, vortices dissipate sooner, limiting their control effect. These time-averaged trends reinforce the phase-averaged observations in Figure~\ref{fig:Q_contour}, confirming that higher duty cycles produce more coherent and persistent vortices.

\subsection{Rapid Lift Estimation for Multivariate Control Testing}
\label{sec:rapid_lift_estimation}
When conducting factorial experiments to optimize SJAs for flow control in various environments, the experimental space can become too large to thoroughly test all possible combinations. Consequently, a single response variable is needed to evaluate the effectiveness of the control strategy. In the absence of force balances, pressure measurements are frequently used to estimate the lift coefficient, though this can be time-consuming. This section aims to correlate single-point pressure measurements with the lift coefficient and assess the feasibility of using such point measurements as the response variable.

\begin{figure}
    \centering
    \begin{subfigure}{0.481\linewidth}
        \includegraphics[width=\linewidth]{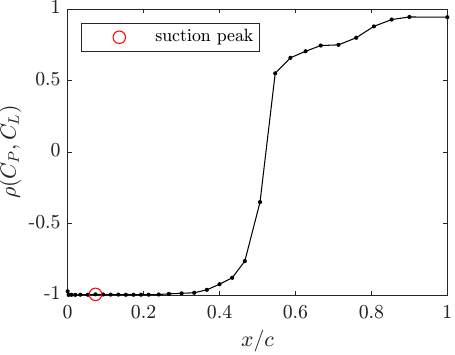}
        \caption{Correlation between the lift coefficient and the pressure coefficient along the chord}
        \label{fig:correlationA}
    \end{subfigure}
    \hfill
    \begin{subfigure}{0.494\linewidth}
        \includegraphics[width=\linewidth]{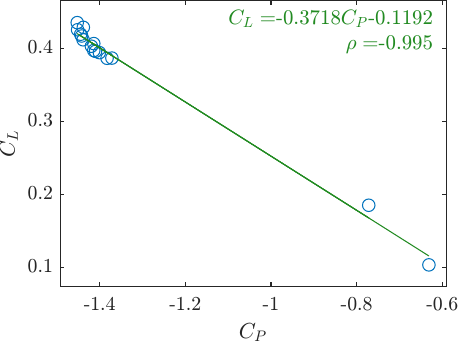}
        \caption{Correlation between the lift coefficient and the pressure coefficient at the suction peak ($x/c=0.07$)}
        \label{fig:correlationB}
    \end{subfigure}
    \caption{Correlations between lift and pressure coefficients using data from all 15 AFC strategies}
    \label{fig:correlation}
\end{figure}

We first investigate the relationship between the overall lift coefficient and the pressure coefficients measured at chordwise locations along the suction surface of the airfoil. The Pearson correlation coefficient, $\rho(C_p, C_L)$, is plotted against the chordwise coordinate in Figure~\ref{fig:correlationA}. This correlation is calculated using data from all 15 control strategies, with each control case providing a single lift coefficient and 33 pressure coefficient measurements from the upper surface of the airfoil. The chordwise location with the highest magnitude of correlation and least variability is found between $x/c = 0.003$ and $0.24$, which includes the suction peak, with $|\rho| \geq 0.995$.

Lift coefficients are plotted against the pressure coefficients of the suction peak in Figure~\ref{fig:correlationB} for the 15 control strategies tested. A strong linear negative correlation is observed, indicating that increased suction pressures at this location are strongly associated with higher lift coefficients. This finding suggests that single-point pressure measurements can serve as effective response variables for both real-time closed-loop flow control, and the optimization of actuation parameters in large factorial experiments using genetic algorithms.

\section{Conclusion}
The effect of duty cycle and blowing ratio on synthetic jet flow control over a stalled NACA 0025 airfoil at $\mathrm{Re}_c=10^5$ and $\alpha=\SI{10}{\degree}$ was studied experimentally. A finite-span microblower array operating at a modulation frequency of $F^+=11.76$ produced varying degrees of flow reattachment under different control strategies. Lift recovery, power efficiency, spanwise control authority, and flow stability were assessed, and the underlying vortical structures responsible for the aerodynamic effects were examined. The combined analysis provides a framework for identifying control strategies that balance aerodynamic performance with power efficiency and flow stability.

The results demonstrated that the threshold momentum coefficient required for flow reattachment can be achieved by increasing either the duty cycle (DC) or the blowing ratio. Control effectiveness increased sharply once flow reattachment was achieved, representing the largest improvement in aerodynamic performance. Additional momentum continued to enhance midspan lift, spanwise control, and flow stability, but the effects eventually saturated, with high-power strategies yielding diminishing returns. The maximum spanwise control length was limited to only \SI{40}{\percent} of the array, highlighting the challenge of extending spanwise control with finite-span SJA arrays. The most power-efficient strategy for time-averaged lift recovery involved low-DCs (\SI{5}{\percent}) at threshold blowing ratios, indicating that brief, strong perturbations can effectively reattach the flow. This approach offers significant power savings compared to conventional strategies using higher duty cycles or continuous actuation.

Low-DC strategies were efficient for time-averaged lift recovery, but flow stability was reduced. At low DCs, SJA-induced vortices were larger, lifted off the surface, and dissipated earlier and inconsistently across phases. This limited momentum transport, producing unsteady flow and, in some cases, a flapping shear layer. In contrast, higher DCs generated smaller, stronger vortices that remained near the surface and maintained more consistent boundary layer control.

Lastly, a strong correlation was observed between the lift coefficient and the strength of the suction peak, demonstrating that single-point pressure measurements can serve as reliable indicators of lift. This finding is particularly important for rapidly testing various control parameters, including modulation frequency, DC, and blowing ratio, to identify optimal combinations for different flow conditions, or to be used in closed-loop flow control systems.

\section*{Funding Sources}
This work was supported by the Natural Science and Engineering Research Council of Canada and the SciNet High Performance Computing consortium and the Canadian Microelectronics Corporation (CMC).

\bibliography{reference}

\end{document}